\documentclass[oneside,sn-nature]{sn-jnl}
\usepackage[T1]{fontenc}
\usepackage[utf8]{inputenc}
\usepackage{colortbl}
\usepackage{booktabs}
\usepackage{bm}
\usepackage{amsmath}
\usepackage{amssymb}
\usepackage{graphicx}
\PassOptionsToPackage{version=3}{mhchem}
\usepackage{mhchem}

\makeatletter

\providecommand{\tabularnewline}{\\}

\usepackage{lipsum}
\newcommand{\vrr}{\vrule width 0pt depth 6pt height 12pt}
\newcommand{\vr}{\vrule width 0pt depth 10pt}

\PassOptionsToPackage{version=3}{mhchem}
\usepackage{mhchem}
\AtBeginDocument{}

\makeatother

\usepackage{babel}
\begin{document}
\title{Chimera states in a network of firefly oscillators}

\author*[1]{\fnm{Deeptarag} \sur{Das}}\email{deeptaragdas@gauhati.ac.in}

\author[2]{\fnm{Debashis} \sur{Saikia}}

\author[1]{\fnm{Kalpana} \sur{Bora}}

\author[1]{\fnm{Madhurjya P.} \sur{Bora}}

\affil[1]{\orgdiv{Physics Department}, \orgname{Gauhati University}, 
\city{Guwahati}  \postcode{781014}, \country{India}}

\affil[2]{\orgdiv{Physics Department}, \orgname{Nabajyoti College}, 
\city{Kalgachia}  \postcode{781319}, \country{India}}

\abstract{We investigate chimera-like states in a network of identical firefly
oscillators with nonlocal delayed coupling. For a uniform communication
delay, the network exhibits a breathing phase-frequency chimera, characterized
by the coexistence of frequency-locked and frequency-drifting populations
and an oscillating incoherent-domain width. We then introduce heterogeneous
pairwise communication delays to investigate a numerical analogue
of the Sarfati-Peleg phenomenon observed in natural firefly swarms.
In this regime, a recurring collective chorus coexists with character
oscillators that flash persistently before or after the chorus. Cycle-resolved
delay maps, role rasters, character probabilities, and spatial covariance
diagnostics demonstrate the persistence and collective organization
of these chorus-character roles. Although the uniform-delay and heterogeneous-delay
states represent different forms of partial synchronization, the same
firefly-oscillator framework supports both regimes. Finally, we construct
a reduced analytical description in terms of a chorus-relative delay
field and a memory field. The local dynamics is organized around a
saturated subcritical pitchfork structure, producing a stable chorus
equilibrium together with two stable character equilibria. The reduced
model therefore provides a phenomenological description of the tri-stable
chorus-character organization and its characteristic delay scale.
Our results show that heterogeneous delayed interactions among identical
relaxation oscillators can generate persistent temporal identities
analogous to those reported in synchronous firefly swarms.}


\maketitle

\section{Introduction}

A chimera state in nonlinear dynamics usually refers to a coexistence
of synchronised (coherent) and unsynchronised (incoherent) populations
in a network of coupled identical nonlinear oscillators \cite{PanaggioAbrams2015}.
Over the years, however, the definition of ‘chimera’ has widened considerably,
and our present analysis addresses one such definition.

Though the original chimera state, as discovered by Kuramoto and Battogtokh
in 2002 \cite{KuramotoBattogtokh2002} (and later named so by Abrams
and Strogatz in 2004 \cite{AbramsStrogatz2004}), was characterised
by the spontaneous coexistence of phase-locked, frequency-synchronised
oscillators and drifting, mutually incoherent oscillators in a network
of identical nonlinear oscillators with non-local (neighbourhood)
coupling, subsequent studies expanded the definition to include global,
all-to-all coupling that can show a self-emergent splitting into synchronous
and asynchronous populations \cite{Yeldesbay2014}. The early 2010s
saw the emergence of the so-called amplitude chimeras \cite{zakharova2014amplitude,bogomolov2016mechanisms},
where the synchronous and disordered groups are related to the amplitude
dynamics of the oscillators. Over the years, we started seeing newer
definitions such as ‘breathing’ and ‘wandering’ chimeras \cite{abrams2008solvable,omelchenko2014capturing}.
In breathing chimeras, the degree of coherence oscillates over time,
while in wandering chimeras the coherent and incoherent regions move
across different nodes of the network. Researchers also reported ‘chimera
death’, when the oscillators do not just lose their alignment; they
stop oscillating, as in oscillation death, and settle into coexisting
spatially coherent and incoherent arrangements of stationary states
\cite{Zakharova2014}. Chimeras have also been investigated in heterogeneous
networks such as networks of model neurons and non-identical oscillators \cite{Laing2009ChimeraHeterogeneous}.
Today, the phrase ‘chimera state’ is often broadly used to describe
self-organised states where coherent and incoherent populations coexist
within a system of coupled oscillators. As we realise, this broad
definition can include a wide variety of phenomena, including the
present investigation.

Like a genie out of a bottle, the idea spread as scientists started
looking for chimera-like behaviour in physical and natural systems,
and the first experimental breakthroughs came in 2012, when two groups
independently demonstrated chimeras in laboratory environments --
one using optoelectronic systems and the other using chemical oscillators
\cite{Hagerstrom2012,Tinsley2012,Martens2013}. To the best of our
knowledge, chimeras, in the broader sense of the term, in naturally
occurring oscillators, such as those in a swarm of fireflies, were
first reported by Sarfati and Peleg in their 2022 paper \cite{SarfatiPeleg2022},
which is remarkable as it presents evidence of a chimera-like phenomenon
in a macroscopic, completely wild and free-roaming animal population.
This is also the subject of the present analysis, as we try to understand
and explain the Sarfati-Peleg (hereafter SP) phenomenon through a
network of numerically modelled fireflies as oscillators. Our primary
assumption of a firefly as a numerical oscillator exhibiting relaxation
oscillations is based on our earlier work on this topic, where we
have been successful in modelling a firefly as a light-emitting chemical
oscillator triggered by the release of a certain chemical (see the
next section) \cite{SaikiaBora2020}. This model successfully explained changes in the periodicity
and amplitude of firefly flashes with respect to ambient temperature.

\subsection{The Sarfati-Peleg phenomenon}

The Sarfati-Peleg (SP) phenomenon refers to the spontaneous coexistence
of distinct flashing groups within a natural swarm of \emph{Photuris
frontalis} fireflies. These groups flash with the same periodicity
but maintain a persistent phase difference, remaining spatially intertwined
\cite{SarfatiPeleg2022}. Sarfati and Peleg interpreted this collective
behaviour as a naturally occurring chimera-like state.

The phenomenon is a chimera-like state as it involves the spontaneous
emergence of distinct dynamical populations within one coupled oscillator
system. It exhibits partial synchronisation as the entire population
of fireflies does not flash or oscillate in unison. The phenomenon
also involves collective organisation without externally assigning
oscillators to different groups. However, we must also note down the
key differences and what makes it distinctive. We note that in a usual
chimera, a phase-locked coherent population coexists with a drifting,
incoherent population, whereas in the SP phenomenon, distinct internally
synchronised groups can coexist with persistent phase differences.
Also groups of fireflies flash with the same periodicity, maintaining
a delay between their flashes, while in the usual chimera, incoherent
oscillators drift relative to the coherent group. Besides, it is important
to note that Sarfati and Peleg observed real fireflies, with individual
variability, movement and an interaction network that was not directly
measured. Coherent and incoherent populations in standard spatial
chimera models often occupy separate regions. In the SP phenomenon,
groups were spatially intertwined. Thus, in case of the SP phenomenon,
the “chimera” is used in a broader sense, extending the concept to
include the coexistence of synchronised groups with distinct collective
phases. The SP state generalizes the original chimera concept by extending
it to include groups of oscillators sharing the same frequency while
maintaining different persistent phases.

\subsection{The numerical experiment}

Given the fact that we have a broad numerical model for the firefly
as a chemical relaxation oscillator, the primary goal of this work
is to see whether we can generate a SP-like phenomenon in a network
of such identical oscillators with non-local coupling in presence
of communication delays. As the analysis shows, we end up using two
time-delay prescriptions -- uniform and heterogeneous time delays.
With uniform time delay, we find coherent and incoherent domains with
different mean phase velocities and drifting relative phases, producing
a breathing phase-frequency chimera. With heterogeneous delay, SP-like
chorus and character oscillator groups appear and have approximately
the same periodicity but maintain persistent early/late phase offsets,
producing the distinctive chorus-character state. So, the former separates
populations mainly by frequency drift and spatial coherence, whereas
the latter mimics the SP-like case by separating the groups of fireflies
mainly by relative flash timing and persistent behavioural roles.

In Section 2, we briefly review the numerical firefly model developed
by Saikia and Bora. In Section 3, we construct a network of identical
firefly oscillators and introduce a communication delay in the coupling
term to account for finite signal-transmission and response times
between neighboring fireflies. In Section 4, we investigate the breathing
phase-frequency chimera exhibited by the network under a uniform time
delay and present the relevant diagnostics supporting its identification.
In Section 5, we introduce heterogeneous pairwise delays, which produce
a Sarfati--Peleg (SP)-like chimera characterized by a persistent
chorus-character state. In Section 6, we discuss the principal signatures
that a reduced analytical model should reproduce. In Section 7, we
present an analytical model based on a saturated subcritical pitchfork
bifurcation. The model exhibits tri-stability, and we argue that the
experimentally observed SP state corresponds to one branch of this
tri-stable equilibrium structure. Finally, in Section 8, we summarise
and conclude.

\section{Firefly as an oscillator}

Our numerical model for a possible chimera state of a network of fireflies
is based on the relaxation oscillator model for fireflies as proposed
by Saikia and Bora in 2020 \cite{SaikiaBora2020}. In this section,
we briefly discuss this mathematical model, which forms the basic
structure of the present analysis.

This model successfully demonstrates that the basic periodic light
emission mechanism of a firefly can be thought to be the result of
a neurological trigger resulting out of accumulation of the chemical
known as D-Luciferin adenylate $(\ce{Luc.LH}_{2}\textrm{-}\ce{AMP})$
inside the peroxisomes (small, membrane-bound cellular organelles
located inside the light-producing cells or photocytes of a firefly's
abdominal lantern). The trigger causes activation of nitric oxide
synthase (NOS) inside the lantern, from which nitric oxide (NO) quickly
diffuses and inhibits binding of oxygen to the photocyte mitochondria
(PM). As a result, oxygen is made available to the peroxisomes for
interaction with the $\ce{Luc.LH}_{2}\textrm{-}\ce{AMP}$ to produce
the oxyluciferin in the excited state $(\ce{Oxyluciferin\ensuremath{^{*}}})$
and an emission occurs \cite{Trimmer2001,SaikiaBora2020}. We note
that Oxyluciferin is the primary chemical product formed when the
light-emitting molecule luciferin is oxidized by the enzyme luciferase
during bioluminescence. 

The fundamental chemical reactions responsible for bioluminescence
in fireflies can be written as 
\begin{eqnarray}
\ce{Luc}+\textrm{D-LH}_{2}+\ce{ATP} & \ce{<=>[\ce{Mg^{2+}}][]} & \ce{Luc.LH_{2}\textrm{-}\ce{AMP}}+\ce{PP_{i}},\label{eq:react1}\\
\ce{Luc.LH_{2}\textrm{-}\ce{AMP}+\ce{O_{2}}} & \ce{->} & \ce{Luc.Oxyluciferin^{*}}+\ce{AMP}+\ce{CO_{2}},\label{eq:react2}\\
\ce{Luc.Oxyluciferin^{*}} & \ce{->} & \ce{Luc.Oxyluciferin}+h\nu.\label{eq:react3}
\end{eqnarray}
In the above reactions, luciferin $(\textrm{D-LH}_{2})$ produces
D-Luciferin adenylate $(\ce{Luc.LH}_{2}\textrm{-}\ce{AMP})$ in presence
of luciferase $(\ce{Luc})$ and adenosine triphosphate $(\ce{ATP})$.
D-Luciferin adenylate, in turn, produces oxyluciferin in the excited
state $(\ce{Oxyluciferin^{*}})$ with the combination of oxygen, which
on de-excitation produces the firefly light $(h\nu)$. The above reactions
can be conveniently expressed as
\begin{alignat}{2}
A & \ce{->} & x, & \qquad(k_{1}),\label{eq:k1}\\
x+y & \ce{->} & B, & \qquad(k_{2}),\label{eq:k2}
\end{alignat}
where $k_{1,2}$ are the corresponding reaction rates with $A$ are
the chain of reactants on the left of the reaction (\ref{eq:react1})
and $B$ represents the products (\ref{eq:react2}) including oxyluciferin.

Saikia and Bora \cite{SaikiaBora2020} proposed that the above model
can be reduced to a two-dimensional oscillator model
\begin{eqnarray}
\frac{dx}{dt} & = & k_{1}A(x)-k_{2}xy,\label{eq:eq1-1}\\
\frac{dy}{dt} & = & \gamma\left(\frac{x}{x_{c}}-1\right)y+\alpha x,\label{eq:2-1}
\end{eqnarray}
where $x$ and $y$, represent the concentration of D-Luciferin adenylate
and oxygen, respectively. The quantity $x_{c}$ is the critical concentration
beyond which the trigger is activated, $\gamma$ is the growth rate
of the instability, and $\alpha$ is a constant. To corresponding
normalized equations can be written as
\begin{eqnarray}
\frac{dx}{dt} & = & x^{2}-\beta xy,\label{eq:dyn1}\\
\frac{dy}{dt} & = & \rho\left(x-1\right)y+\eta x,\label{eq:dyn2}
\end{eqnarray}
where the following scaling and normalization are used
\begin{align*}
t & \to tk_{1}, & x & \to xx_{c}, &  & A(x)\to x_{c}A(x),\\
\rho & =\frac{\gamma}{k_{1}}, & \eta & =\frac{\alpha x_{c}}{k_{1}}, &  & \beta=\frac{k_{2}}{k_{1}},
\end{align*}
with the constraints $\beta,\rho,\eta>0$. Eqs.(\ref{eq:dyn1},\ref{eq:dyn2})
are shown to have a stable limit cycle, which demonstrates a relaxation
oscillation cycle. In the above equations, the parameter $A(x)$ is
set to $x^{2}$. This system has been proven to provide a convincing
firefly flash model in the experimentally demonstrated parameter regime.
In the rest of our analysis, we fix the oscillator parameters as $\beta=0.052$,
$\rho=1.8$, and $\eta=8.2$, corresponding to firefly flashing cycle
at $20\,^{\circ}\textrm{C}$, which has a normalized period $T_{0}\simeq14.43$.

\section{Network of oscillators and coupling}

We now consider a network of $N$ identical firefly oscillators as
represented by Eqs.(\ref{eq:dyn1},\ref{eq:dyn2}). This ensemble
of oscillators form a closed oscillator system, which can be written
with a periodic indexing as,
\begin{equation}
(x_{i},y_{i})\,\textrm{for}\,i=1,2,\dots,N,\quad\textrm{with}\,(x_{0},y_{0})=(x_{N},y_{N}),\label{eq:couple1}
\end{equation}
so that $(x_{N+1},y_{N+1})=(x_{1},y_{1})$. In general 
\begin{equation}
(x,y)_{i+m}=(x,y)_{1+[(i+m-1)\,\textrm{mod}\,N]}.
\end{equation}
The coupling between individual oscillators is achieved through a
coupling term in the second oscillator equation 
\begin{eqnarray}
\frac{dx_{i}}{dt} & = & x^{2}_{i}-\beta x_{i}y_{i},\\
\frac{dy_{i}}{dt} & = & \eta x_{i}+\rho(x_{i}-1)y_{i}+\kappa{\cal C}_{i},\label{eq:couple4}
\end{eqnarray}
where $\kappa$ is the coupling strength and ${\cal C}_{i}$ is the
coupling term. The coupling is introduced into the second equation
because $y_{i}$, representing the oxygen concentration, directly
controls the rapid excitation associated with light emission, whereas
$x_{i}$ describes the slower accumulation of D-Luciferin adenylate.
Thus, the flashes of neighbouring fireflies most naturally influence
the excitation variable $y_{i}$. Any delay in this coupling may be
interpreted as a communication delay -- the finite time required
for a firefly to perceive, process, and respond to the optical signal
emitted by another firefly \cite{SchusterWagner1989,Sethia2008}.

\subsection{Ring ($R$ nearest neighbours) coupling}

In our network of oscillators, we use a ring coupling with $R$ nearest
neighbours, which can be mathematically represented as 
\begin{equation}
\frac{dy_{i}}{dt}=\eta x_{i}+\rho_{1}(x_{i}-1)y_{i}+\frac{\kappa}{2R}\sum^{R}_{m=1}[(y_{i+m}-y_{i})+(y_{i-m}-y_{i})],\label{eq:ring1}
\end{equation}
with 
\begin{equation}
1\leq R\leq\left\lfloor N/2\right\rfloor .\label{eq:ring2}
\end{equation}
or equivalently 
\begin{equation}
\frac{dy_{i}}{dt}=\eta x_{i}+\rho_{1}(x_{i}-1)y_{i}+\frac{\kappa}{2R}\sum^{R}_{m=1}(y_{i+m}+y_{i-m}-2y_{i})],\label{eq:ring3}
\end{equation}
where $R$ is known as the `coupling radius'. This coupling reduces
to the nearest-neighbour coupling for $R=1$.

We note that natural causes in case of a swarm of fireflies in the
forest which can affect the response of individual fireflies, each
firefly responds primarily to nearby visible neighbours rather than
to the entire population. To this effect, ring coupling provides an
idealized representation of the spatially local communication among
fireflies in its natural environment. Arranging the oscillators on
a ring imposes periodic boundaries and avoids artificial edge effects,
while nonlocal ring coupling allows each oscillator to interact with
a finite neighbourhood on either side. Thus, a ring coupling is a
minimal model that captures local, distance-limited visual communication
in a spatially extended firefly population.

\subsection{Uniform time delay}

After one firefly emits a flash, the signal must be detected and processed
by the sensory and neural system of another firefly before it modifies
the biochemical excitation controlling the next flash. Although the
propagation time of light is negligible, this perception-response
time is finite and to model this perception-response delay, we introduce
a uniform time delay in the coupling term, which represents the finite
communication and response time between the fireflies. Additional
effective delays may arise from intermittent visibility and obstruction
by other natural causes and hence, the state of a neighbouring firefly
at an earlier time, $(t-\tau)$, rather than at the instantaneous
time $t$, must be introduced in the coupling term with $\tau$ as
the time delay. 

With time delay $\tau$, the oscillator system becomes 
\begin{eqnarray}
\dot{x}_{i}(t) & = & x_{i}(t)^{2}-\beta x_{i}(t)\,y_{i}(t),\label{eq:ring-1}\\
\dot{y}_{i}(t) & = & \eta x_{i}(t)+\rho_{1}[x_{i}(t)-1]y_{i}(t)\nonumber \\
 &  & +\,\frac{\kappa}{2R}\sum^{R}_{m=1}[y_{i+m}(t-\tau)+y_{i-m}(t-\tau)-2y_{i}(t)].\label{eq:ring-2}
\end{eqnarray}

\section{Search for a chimera state}

As we know, an usual chimera state is the coexistence of synchronized
and desynchronized groups in a network of $N$ identical oscillators.
We now search for such a state in our network using the ring-type
coupling described earlier \cite{KuramotoBattogtokh2002,Omelchenko2011}.

We start with a periodic indexing i.e. 
\begin{equation}
i+N\equiv i,
\end{equation}
with $N=100$, which represents a closed ring of oscillators. We also
set 
\begin{equation}
R=rN,
\end{equation}
where $r$ is the fraction of the neighbouring oscillators which are
coupled for a given oscillator. A  chimera state requires intermediate
nonlocal coupling in which each oscillator interacts with enough neighbours
to form coherent domains, but not so many that global synchronization
suppresses the incoherent domain. We therefore choose a value of $r\sim0.15-0.35$
for this search.

\begin{figure}
\begin{centering}
\includegraphics[width=0.49\textwidth]{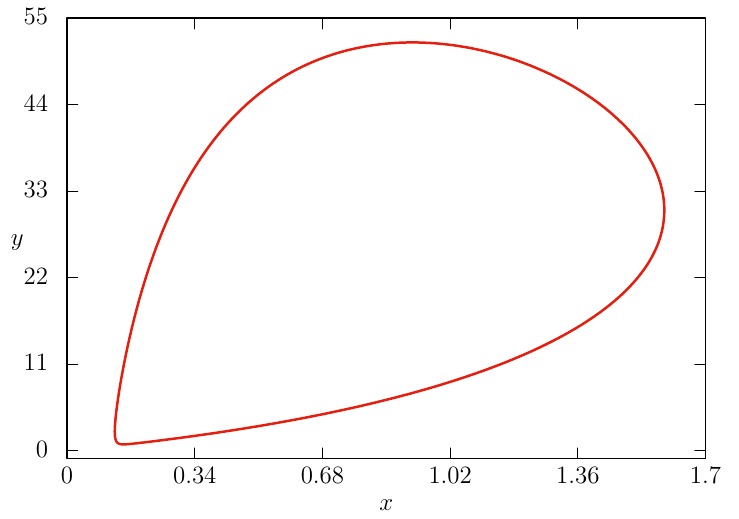}\hfill{}\includegraphics[width=0.49\textwidth]{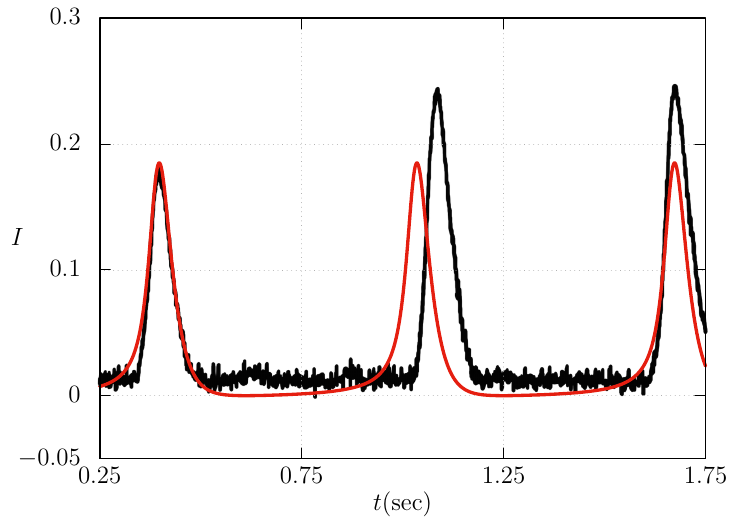}
\par\end{centering}
\caption{The limit cycle of a single oscillator (in normalized units) is shown
in the left panel and its mapping with the corresponding experimental
data at $20\,^{\circ}\textrm{C}$ (in real physical units) is shown
in the right panel (from \cite{SaikiaBora2020}).}\label{fig:The-limit-cycle}
\end{figure}

\subsection{Initial conditions}

We start from the stable limit cycle of a single oscillator, which
is shown in Fig.\ref{fig:The-limit-cycle}. Assuming that $(x_{0},y_{0})$
is a point on the stable limit cycle of the oscillator, we use a nonuniform
initial conditions for the ring oscillator system, around $(x_{0},y_{0})$,
\begin{eqnarray}
x_{i}(0) & = & x_{0}(\phi_{i})+\epsilon_{i},\\
y_{i}(0) & = & y_{0}(\phi_{i})+\delta_{i},
\end{eqnarray}
with 
\begin{equation}
\phi_{i}=\phi_{0}+A\,\cos\left(\frac{2\pi i}{N}\right),
\end{equation}
where $\phi_{0}$ is an initial arbitrary phase, which can also be
zero. In the above expression, $A$ is the amplitude of spatial modulations
and $(\epsilon,\delta)_{i}$ are random perturbations for the $i$th
oscillator. Note that $(\epsilon,\delta)_{i}$ are different for each
oscillator, as these modulations and perturbations are necessary for
a chimera seeding. In our numerical search, we have set $A=0.4$,
with $\epsilon_{i}$ and $\delta_{i}$ independently sampled from
a uniform distribution over $[-10^{-3},10^{-3}]$.

\subsubsection{Parameter scanning}

We start our parameter scanning in the $(\kappa,\tau)$ space. While
very low values of $\kappa$ (such as $0.001$) provide a very weak
coupling which can mostly lead to incoherence, sufficiently large
values may lead to a synchronous state. The crossover to synchrony
is not fixed, but depends on the communication delay and coupling
topology. So, we look for intermediate values for $\kappa$ for a
possible chimera state. We note that while there is no standard prescription
for intermediate values, we refine the scan around the transition
between incoherence and synchrony; the candidate cases retained below
have $0.16\leq\kappa\leq0.17$. For the time delay, we set 
\begin{equation}
\tau=\nu T_{0},\quad\nu\in\{0.05,0.1,0.2,0.5,\dots\},
\end{equation}
where $T_{0}$ is the natural period of the limit cycle. This relation
is used only to set the natural delay scale and to guide the initial
coarse parameter scan. We subsequently scan $\tau$ directly over
a wider range and refine the sampling around the most promising candidates.
 In the broad scanning of the parameter space, we use the following
diagnostic signatures.
\begin{itemize}
\item \textbf{Space-time diagram}\\
After integrating the system for long enough time for transients to
die down, the space-time plot for the variable $y_{i}(t)$ is plotted.
While a smooth colour map indicates a synchronous state, a random
scattered plot, then it points toward a complete incoherent state.
For a possible chimera state, the plot should be a generous mix of
smooth and incoherent domains.
\item \textbf{Instantaneous phase}\\
As ours is a system of real oscillators rather than phase oscillators
and we \emph{must} construct the instantaneous phase $\theta_{i}(t)$
for each oscillator. For this purpose, we construct the geometric
phase for each oscillator 
\begin{equation}
\theta_{i}(t)=\textrm{arctan2}(y_{i}(t)-\bar{y},x_{i}(t)-\bar{x}),
\end{equation}
where $(\bar{x},\bar{y})$ is the `centre' of the limit cycle. In
the above expression 
\begin{equation}
\textrm{arctan2}(y,x)=\left\{ \begin{array}{ll}
\tan^{-1}(y/x), & x>0\\
\tan^{-1}(y/x)+\pi, & x<0,y\geq0\\
\tan^{-1}(y/x)-\pi, & x<0,y<0\\
\pi/2, & x=0,y>0\\
-\pi/2 & x=0,y<0\\
\textrm{undefined}, & x=y=0
\end{array}\right.
\end{equation}
We also compute the phase differences 
\begin{equation}
\Delta_{i}(t)=\arg\left[e^{i(\theta_{i+1}-\theta_{i})}\right].
\end{equation}
In the case of a chimera state, $\Delta_{i}$ should vary smoothly
in the coherent regions and should vary wildly (or irregularly) in
incoherent regions. Mathematically 
\begin{equation}
\bar{x}=\frac{1}{T_{0}}\int^{t_{0}+T_{0}}_{t_{0}}x(t)\,dt,\quad\bar{y}=\frac{1}{T_{0}}\int^{t_{0}+T_{0}}_{t_{0}}y(t)\,dt,
\end{equation}
or time-average over one period of the limit cycle, where $T_{0}$
is the period of the limit cycle. Numerically however, we adopt a
practical way to find it instead of integrating by putting the exact
value of $T_{0}$. So, the procedure can be outlined as follows:
\begin{enumerate}
\item Integrate a single oscillator without coupling for a long enough time
so that it is \emph{fully} on the limit cycle.
\item Take enough time steps (for the limit cycle to be completed several
times) and compute the following averages 
\begin{equation}
\bar{x}=\frac{1}{N_{t}}\sum^{N_{t}}_{n=1}x(t_{n}),\quad\bar{y}=\frac{1}{N_{t}}\sum^{N_{t}}_{n=1}y(t_{n}),
\end{equation}
where $N_{t}$ is the total number of time steps. 
\end{enumerate}
\item \textbf{Order parameter}\\
The local order parameter $Z_{i}(t)$ \cite{KuramotoBattogtokh2002,Omelchenko2011,zakharova2014amplitude}
provides another diagnostics
\begin{equation}
Z_{i}(t)=\left|\frac{1}{(2\delta+1)}\sum_{|j-i|\leq\delta}e^{i\theta_{j}(t)}\right|,
\end{equation}
where $\delta$ is a small local averaging window, typically $2\leq\delta\leq10$.
In the present calculation, we use $\delta=3$, corresponding to an
average over $2\delta+1=7$ neighbouring sites including oscillator
$i$. We have the following outcomes: 
\begin{align}
Z_{i}(t)\approx1, & \quad\textrm{locally coherent},\\
Z_{i}(t)<1, & \quad\textrm{locally incoherent}.
\end{align}
For a chimera state, there should be both the above outcomes in separate
domains.  For a traveling chimera, the domains keep on moving in time.
For breathing chimera, the size of the domains may oscillate in time.
\end{itemize}
Based on the above diagnostic check, we have isolated at least three
possible chimera state as listed in Table.\ref{tab:Parameters-for-possible}

\begin{table}
\caption{Parameters for possible chimera regime. The case indicated by `$\protect\dag$'
is the one for which the diagnostics are presented.\vr}\label{tab:Parameters-for-possible}

\centering{}%
\begin{tabular}{|c|c|}
\hline 
\textbf{Parameters} \vrr & \textbf{Value(s)}\tabularnewline
\hline 
\hline 
$N$ \vrr & $100$\tabularnewline
\hline 
$R$ \vrr & $16$\tabularnewline
\hline 
$(\kappa,\tau)$ \vrr & $(0.17,1.25)^{\dag},(0.17,1.5),(0.16,1.25)$\tabularnewline
\hline 
$\delta$ \vrr & $3$\tabularnewline
\hline 
\end{tabular}
\end{table}

\subsection{Distinguishing the chimera state}

In a system of oscillators, one \emph{must} distinguish a cluster
state from a chimera state. While a cluster state may have several
differently synchronised groups or domains, but each group is internally
coherent. So, inspection of $\theta_{i}(t)$ and $\Delta_{i}(t)$
is crucial for this distinction. We note that for a fully synchronous
system we should observe a nearly flat line in $\theta_{i}$ and zero
$\Delta_{i}$ 
\begin{equation}
\theta_{i}\approx\textrm{constant},\quad\Delta_{i}\approx0\,\forall i.
\end{equation}
For a phase-locked traveling wave,
\begin{equation}
\begin{array}{rcl}
\theta_{i} & \approx & qi+\textrm{constant},\\
\Delta_{i} & \approx & q\,(\textrm{same}\,\forall i).
\end{array}
\end{equation}
Yet another possible observation may be cluster state, which can be
difficult to distinguish from a true chimera state. In this case,
the oscillators split into a few or several clusters and each cluster
is synchronized. So, we should probably have 
\begin{equation}
\begin{array}{rl}
\theta_{i} & :\quad\textrm{piecewise constant}\\
\Delta_{i} & :\quad\approx0\,\textrm{everywhere, with isolated discontinuities}.
\end{array}
\end{equation}
Besides, for a cluster state, we should also have 
\begin{equation}
Z_{i}\approx1\,\textrm{everywhere}.
\end{equation}
For a true chimera state, we should however have 
\begin{equation}
\begin{array}{rl}
\theta_{i} & :\quad\textrm{smooth in one domain, irregular in another},\\
\Delta_{i} & :\quad\approx\,\textrm{constant or slowly varying}\:(\textrm{in coherent regions}),\\
 & :\quad\textrm{strongly fluctuating / irregular \ }(\textrm{in incoherent regions}).
\end{array}
\end{equation}
Besides, a chimera-state signature should be robust under small changes
in the parameters and initial conditions. In the present study, we
keep $N=100$ fixed and examine robustness under changes in the coupling
range $R$ (and hence in $R/N$), nearby values of $\kappa$ and $\tau$,
random initial-condition seeds, and longer integration times. The
signatures should not vanish under these small variations.

\subsection{Diagnostics and results}

Among three best cases isolated for a chimera state, as shown in Table.\ref{tab:Parameters-for-possible},
we present the diagnostics for the case $(\kappa,\tau)=(0.17,1.25)$
indicated by `$\dag$' in the table.

As for the diagnostics, we shall use the following quantities, as
described in the subsequent subsections. We also carry out these diagnostics
in a late-time interval $[t_{0},t_{1}]=[200,400]$ so as to confirm
the persistence of the chimera state. This late-time interval is further
divided into four late-time windows $W_{m}$ with 
\begin{equation}
W_{1}=[200,250],\quad W_{2}=[250,300],\quad W_{3}=[300,350],\quad W_{4}=[350,400].
\end{equation}
We note that the phase $\theta_{i}(t)\in(-\pi,\pi]$ jumps by $2\pi$
whenever an oscillator completes a cycle. To overcome these jumps,
we define the unwrapped phase $\Theta_{i}(t)$ as 
\begin{equation}
\Theta_{i}(t)=\theta_{i}(t)+2\pi k_{i}(t),
\end{equation}
where $k_{i}(t)$ changes by an integer whenever such jumps occurs.
Numerically, if the sampled time is $t_{n}$, then 
\begin{equation}
k_{i}(t_{n})=\sum^{n}_{\ell=1}q_{i}(t_{\ell}),
\end{equation}
where 
\begin{equation}
q_{i}(t_{\ell})=\left\{ \begin{array}{rl}
+1, & \theta_{i}(t_{\ell})-\theta_{i}(t_{\ell-1})<-\pi,\\
-1, & \theta_{i}(t_{\ell})-\theta_{i}(t_{\ell-1})>+\pi,\\
0, & \textrm{otherwise.}
\end{array}\right.
\end{equation}

\subsubsection{Mean phase velocity}

We define the mean phase velocity $\bar{\omega}_{i}$ for each oscillator
in the late-time as 
\begin{equation}
\bar{\omega}_{i}=\frac{\Theta_{i}(t_{1})-\Theta_{i}(t_{0})}{t_{1}-t_{0}},\quad i=1,2,\dots,N.
\end{equation}
Accordingly, the windowed mean phase velocity $\bar{\omega}^{(m)}_{i}$
is defined as 
\begin{equation}
\bar{\omega}^{(m)}_{i}=\frac{\Theta_{i}(t_{m+1})-\Theta_{i}(t_{m})}{t_{m+1}-t_{m}},\quad i=1,2,\dots,N,\quad m=1,2,3,4.
\end{equation}

\subsubsection{Mean local order parameter}

Accordingly, we define the mean local order parameter $\bar{Z}_{i}$
as 
\begin{equation}
\bar{Z}_{i}=\frac{1}{t_{1}-t_{0}}\int^{t_{1}}_{t_{0}}Z_{i}(t)\,dt,\quad i=1,2,\dots,N.
\end{equation}
For our sampled discrete data, we have equivalently, 
\begin{equation}
\bar{Z}_{i}=\frac{1}{N_{I}}\sum_{t_{k}\in I}Z_{i}(t_{k}),\quad I=[t_{0},t_{1}],
\end{equation}
where $I=[t_{0},t_{1}]$ contains the discrete numerical intervals.
Subsequently, the windowed local coherence $\bar{Z}^{(m)}_{i}$ is
defined as 
\begin{equation}
\bar{Z}^{(m)}_{i}=\frac{1}{t_{m+1}-t_{m}}\int^{t_{m+1}}_{t_{m}}Z_{i}(t)\,dt,\quad i=1,2,\dots,N,\quad m=1,2,3,4.
\end{equation}
For the sampled data, we have 
\begin{equation}
\bar{Z}^{(m)}_{i}=\frac{1}{N_{m}}\sum_{t_{k}\in W_{m}}Z_{i}(t_{k}),
\end{equation}
where $N_{m}$ is the number of sampled time points for window $W_{m}$,
which is $4$ in our case.

\begin{figure}
\begin{centering}
\includegraphics[width=0.49\textwidth]{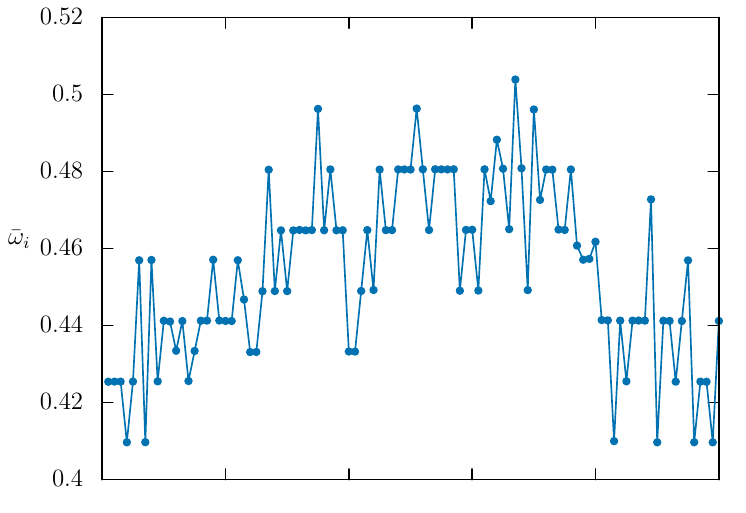}\hfill{}\includegraphics[width=0.495\textwidth,totalheight=1.78in]{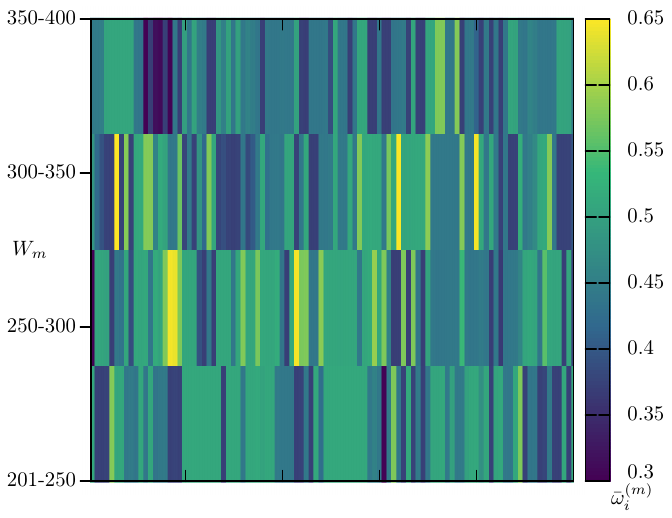}\\
 \includegraphics[width=0.49\textwidth]{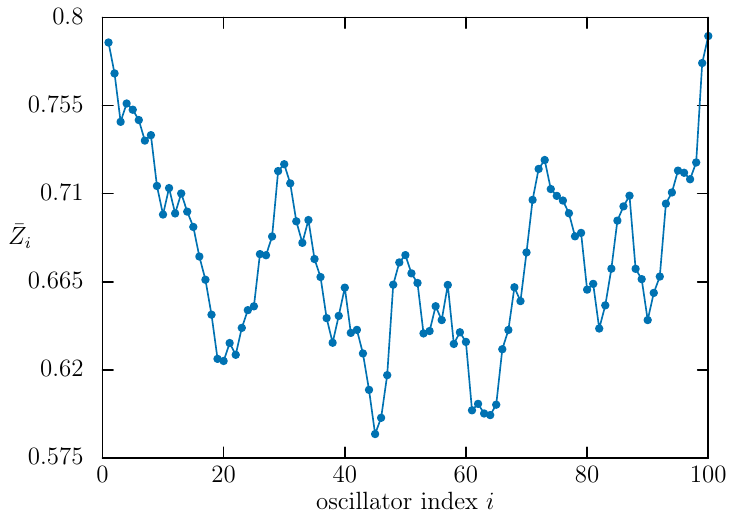}\hfill{}\includegraphics[width=0.495\textwidth,totalheight=1.78in]{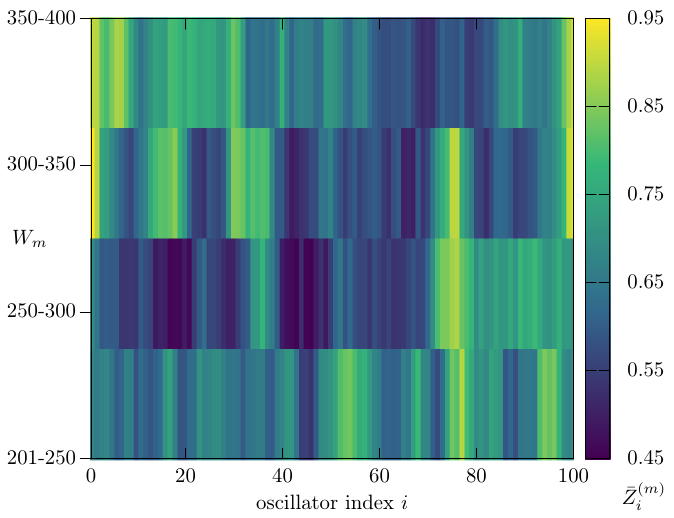}
\par\end{centering}
\caption{The late-time mean phase velocity $\bar{\omega}_{i}$ and mean local
order parameter $\bar{Z}_{i}$ are shown in the left-hand column.
The corresponding windowed mean phase velocity $\bar{\omega}^{(m)}_{i}$
and windowed local coherence $\bar{Z}^{(m)}_{i}$ are shown in the
right hand column.}\label{fig:The-late-time-mean}
\end{figure}

Fig.\ref{fig:The-late-time-mean} plots these four quantities. The
mean phase velocity $\bar{\omega}_{i}$ clearly shows the coexistence
of frequency-locked and frequency-drifting populations. Similarly,
the mean local order parameter $\bar{Z}_{i}$ shows both high- (brighter)
and low-coherence (darker) regions around the same ring of oscillators
supporting the coexistence of coherent and incoherent domains at the
same time. The windowed mean phase velocity $\bar{\omega}^{(m)}_{i}$
shows spatial variation in the average phase velocity. As can be seen,
the pattern is not transient that disappears after one interval but
a heterogeneous frequency structure does persist. The shifting bright/dark
bands suggests a breathing or wandering chimera rather than a perfectly
stationary one. This is also supported by the windowed local coherence
$\bar{Z}^{(m)}_{i}$, as each window contains both coherent regions
(brighter) and less coherent regions (darker). The persistence of
mixed high/low coherence supports robustness of the chimera-like split.
The changing location/intensity of the bands also shows that the coherent
and incoherent groups fluctuate, but the coexistence remains.

\subsubsection{Space-time plot}

\begin{figure}
\begin{centering}
\includegraphics[width=0.5\textwidth]{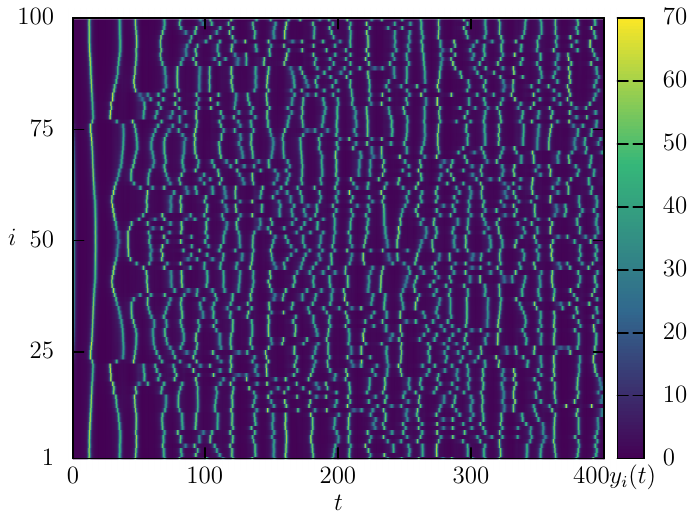}
\par\end{centering}
\caption{Space-time plot of the state variable $y_{i}(t)$ in time $t$ over
oscillator index $i$.}\label{fig:Space-time-plot-of}
\end{figure}

As a primary visualization, we present the space-time plot for the
oscillator state variable $y_{i}(t)$ over oscillator index $i$ in
time $t$. This is shown in Fig.\ref{fig:Space-time-plot-of}, where
we see coexistence of almost vertical lines with zagged lines indicating
the existence of synchronized firings as well as desynchronized firings
throughout the time domain except in the initial transient region.

\subsection{A breathing chimera?}

\begin{figure}
\begin{centering}
\includegraphics[width=0.9\textwidth]{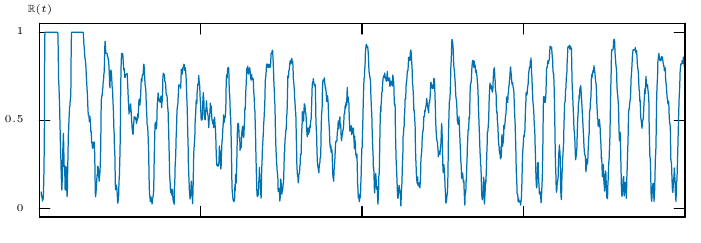}\\
 \vskip-10pt\includegraphics[width=0.89\textwidth]{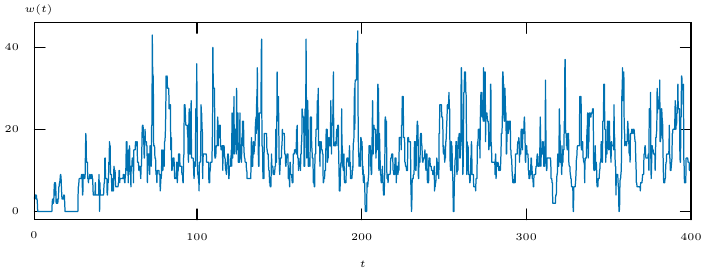}
\par\end{centering}
\caption{The behaviours of the global order parameter $\mathbb{R}(t)$ and
incoherent width $w(t)$.}\label{fig:The-behaviour-of}
\end{figure}

\begin{figure}
\begin{centering}
\includegraphics[width=0.5\textwidth]{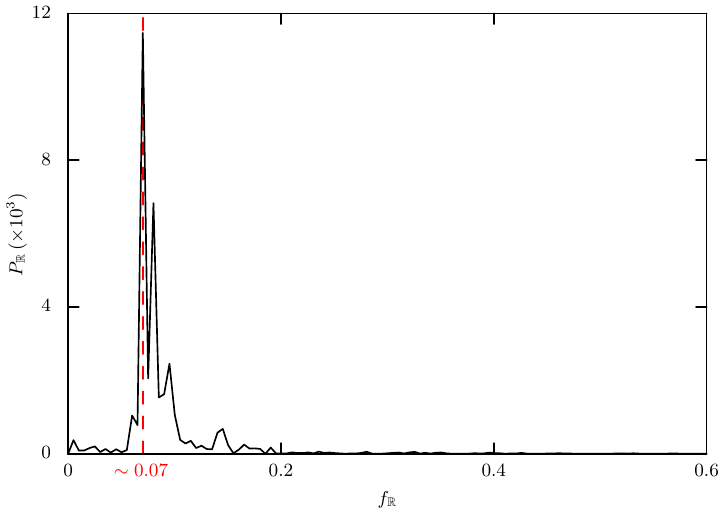}\hfill{}\includegraphics[width=0.5\textwidth]{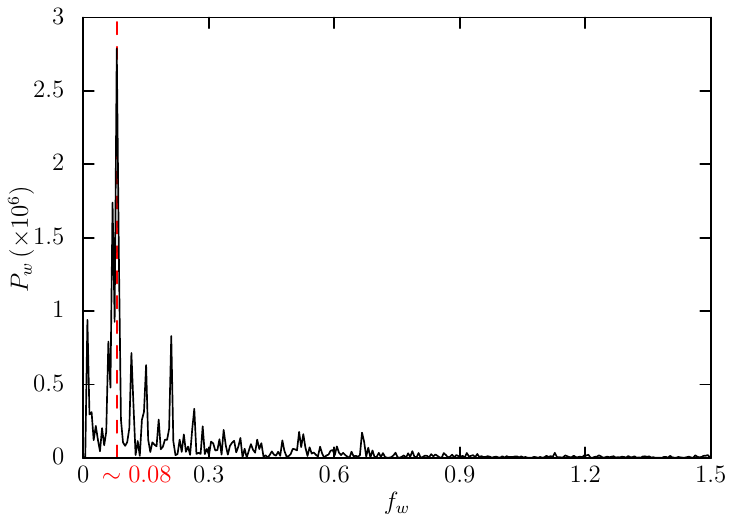}
\par\end{centering}
\caption{Power spectra $P_{\mathbb{R},w}$ of the global order parameter profile
and the incoherent width profile.}\label{fig:Power-spectra-}
\end{figure}

We now numerically show that the chimera state we have just described
is probably a breathing chimera \cite{Zakharova2014}. In support
of this claim, we present the global order parameter $\mathbb{R}(t)$,
\begin{equation}
\mathbb{R}(t)=\left|\frac{1}{N}\sum^{N}_{i=1}e^{i\theta_{i}(t)}\right|.
\end{equation}
While for a strongly phase-synchronized population of oscillators,
we expect $\mathbb{R}\approx1$, our case shows a very well-defined
oscillatory behaviour (see Fig.\ref{fig:The-behaviour-of} top panel)
with a dominant frequency at $\sim0.07$ (see Fig.\ref{fig:Power-spectra-},
left panel), as shown in the power spectrum. This clearly shows that
the chimera state is a breathing one. This is also supported by the
\emph{incoherent domain width} $w(t)$, defined as, 
\begin{equation}
w(t)=\sum^{N}_{i=1}\mathbf{1}[Z_{i}(t)<Z_{\textrm{th}}],
\end{equation}
where $Z_{\textrm{th}}$ is the threshold local order parameter (or
local coherence), which is determined empirically from the local-coherence
field of the flagship run. Specifically, for $\delta=3$, we define
$Z_{\textrm{th}}=Q_{0.35}\left(\{Z_{i}(t_{k})\}_{i,k}\right)$, where
$Q_{0.35}$ denotes the $35$th percentile over all oscillator indices
and sampled times. This selects the lower-coherence part of the empirical
distribution and gives $Z_{\textrm{th}}=0.4573$ for the present data
set. Thus, this is an operational, data-dependent threshold rather
than a universal constant. The quantity $\mathbf{1}[\cdot]$ is $1$
if the condition is true and $0$ if it is false. From the late-time
mean-phase-velocity profile, the representative mean phase velocity
of the system is evaluated as $\bar{\omega}\simeq0.45$, corresponding
to the mean phase frequency $f_{\bar{\omega}}=\bar{\omega}/(2\pi)\simeq0.072$.
The behaviour of $w(t)$ is also shown in the bottom panel of Fig.\ref{fig:The-behaviour-of}
and the corresponding power spectrum is shown in the right panel of
Fig.\ref{fig:Power-spectra-} The fact that width $w(t)$ also shows
an oscillatory behaviour with a dominant frequency at $\sim0.08$,
indicates that the global coherence modulation and the expansion (or
contraction) of the incoherent domain are part of the same breathing
process. Interestingly, this mean phase frequency is close to the
dominant frequencies $f_{\mathbb{R},w}$, suggesting that the breathing
process is likely locked to the natural flashing (or relaxation) cycle
of the oscillators.

\section{A Sarfati-Peleg-like chimera state}

To investigate a possible chimera state with reference to the SP phenomenon,
we use a \emph{heterogeneous pairwise communication delay} to the
ring coupling, to investigate whether delay heterogeneity can produce
the chorus-character organisation absent from the uniform-delay case.

The motivation for this modification stems from the experimental observations
of Sarfati and Peleg \cite{SarfatiPeleg2022}, where individual fireflies
respond to collective flashes with different response latencies, giving
rise to the coexistence of ``chorus'' and ``character'' flashes.
Rather than prescribing different oscillator properties, we assume
that all fireflies are intrinsically identical and attribute the observed
temporal differentiation entirely to the interaction network. We note
that an SP-like chimera is organized by flashing roles rather than
only by instantaneous phase coherence. A dominant group produces synchronized
chorus flashes, while persistent character fireflies flash systematically
before or after the chorus and may switch roles over time. Thus, we
propose that the essential signature is a structured, persistent distribution
of flash delays and behavioural identities, generated by heterogeneous
communication delays, which differs from a mere spatial division into
coherent and incoherent domains.

\subsection{The delay topology}

As before, we use a periodic indexing as given by Eqs.(\ref{eq:couple1}-\ref{eq:couple4})
with ring coupling defined by Eqs.(\ref{eq:ring1}-\ref{eq:ring3}).
In our implementation, we use a `ring-uniform kernel', which is defined
by a neighbourhood ${\cal N}_{i}$ of the $i$th oscillator, defined
as 
\begin{equation}
{\cal N}_{i}=\{i-R,\dots,i-1,i+1,\dots,i+R\},
\end{equation}
so that for any ${\cal N}_{i}$, we have $|{\cal N}_{i}|=2R$. As
we use a heterogeneous pairwise delay, we introduce a pairwise delay
$\tau_{ij}$ and the coupling term becomes 
\begin{equation}
{\cal C}_{i}=\sum_{j\in{\cal N}_{i}}w_{ij}[y_{j}(t-\tau_{ij})-y_{i}(t)],
\end{equation}
where $w_{ij}$ are coupling weights which satisfy the relation 
\begin{equation}
w_{ij}=\frac{1}{2R},\quad j\in{\cal N}_{i},
\end{equation}
so that 
\begin{equation}
\sum_{j\in{\cal N}_{i}}w_{ij}=1.
\end{equation}
For introducing the heterogeneity, we define the following relation
for each interacting pair 
\begin{equation}
\tau_{ij}=\tau_{0}+\varrho\frac{d_{ij}}{R}+\sigma_{\tau}\xi_{ij},
\end{equation}
where $\tau_{0}$ is the mean communication delay and $d_{ij}$ is
the shortest ring distance given by 
\begin{equation}
d_{ij}=\min(|i-j|,N-|i-j|).
\end{equation}
The pairwise delay consists of three parts -- the mean delay $\tau_{0}$,
a distance-dependent delay (the $d_{ij}/R$ term), and a random heterogeneous
delay (the $\sigma_{\tau}$ term). The quantity $\sigma_{\tau}$ determines
the magnitude of the delay heterogeneity and $\xi_{ij}$ is an independent
uniformly distributed random number 
\begin{equation}
\xi_{ij}\in[-1,1].
\end{equation}
Accordingly, 
\begin{equation}
-\sigma_{\tau}\leq\sigma_{\tau}\xi_{ij}\leq+\sigma_{\tau},
\end{equation}
so that the delays fluctuate about the mean value $\tau_{0}$ within
an interval determined by $\sigma_{\tau}$. The term $\varrho(d_{ij}/R)$
introduces a weak deterministic increase of the delay with a spatial
separation. This separation is bounded between 
\begin{equation}
0\leq\varrho\frac{d_{ij}}{R}\leq\varrho
\end{equation}
and it acts as a small bias relative to the random delay component.
It also gives us a numerical leverage for fine-scale tweaking of the
delayed network. For numerical consistency, we \emph{must }have 
\begin{equation}
\tau_{ij}\geq\Delta t,
\end{equation}
where $\Delta t$ is the numerical integration time step. In our simulations,
we have used $\varrho=0.015$, though a similar small value would
also have been sufficient.

\begin{table}
\caption{Parameters for Sarfati-Peleg-like chimera state \cite{SarfatiPeleg2022}.
This table lists top $8$ best candidates.\vr}\label{tab:Parameters-for-Sarfati-Peleg-lik}

\centering{}%
\begin{tabular}{|c|c|c|c|c|c|}
\hline 
\multicolumn{6}{|c|}{\textbf{Parameters and Values}\vrr}\tabularnewline
\hline 
\textbf{Sl. No.} & $N$ & $R$ & $\kappa$ & $\tau_{0}$ & $\sigma_{\tau}$\vrr\tabularnewline
\hline 
\hline 
1 \vrr & $100$ & $9$ & $0.0485$ & $1.185$ & $0.062$\tabularnewline
\hline 
2 \vrr & $100$ & $9$ & $0.0486$ & $1.19$ & $0.052$\tabularnewline
\hline 
3 \vrr & $100$ & $9$ & $0.0486$ & $1.19$ & $0.051$\tabularnewline
\hline 
4 \vrr & $100$ & $9$ & $0.0485$ & $1.185$ & $0.06$\tabularnewline
\hline 
5 \vrr & $100$ & $9$ & $0.0485$ & $1.185$ & $0.061$\tabularnewline
\hline 
6 \vrr & $100$ & $9$ & $0.0486$ & $1.19$ & $0.05$\tabularnewline
\hline 
7 \vrr & $100$ & $9$ & $0.0485$ & $1.185$ & $0.063$\tabularnewline
\hline 
8 \vrr & $100$ & $9$ & $0.0485$ & $1.187$ & $0.069$\tabularnewline
\hline 
\end{tabular}
\end{table}

Physically, the distance-dependent delay mimics the delay associated
with spatial separation of the fireflies. This can be justified in
the respect of delay due to longer viewing distance, reduced visibility,
and longer effective reaction because distant flashes are detected
less promptly. It is a deterministic geometric correction and is very
small compared with $\tau_{0}$. In terms physical swarm of fireflies,
the random variability term represents individual or behavioural variability,
and environmental effects. This term has no preferred spatial direction,
it simply makes every interaction slightly different.

Finally, the coupling term can be written as 
\begin{equation}
{\cal C}_{i}=\frac{1}{2R}\sum^{R}_{m=1}[y_{i+m}(t-\tau_{i,i+m})+y_{i-m}(t-\tau_{i,i-m})-2y_{i}(t)].
\end{equation}
For constant delay $\tau_{ij}=\tau$ and the above coupling term reduces
to the one used in Eq.(\ref{eq:ring-2}).

\subsection{Diagnostics for SP-like effect}

As a numerical proof, we present the following Tier-I diagnostics: 
\begin{enumerate}
\item Collective flash count and chorus, 
\item Cycle-resolved delay, 
\item Role raster, and 
\item Character probability 
\end{enumerate}
We have identified several parameter regimes where the SP-like effect
is quite pronounced. Out of these, Table.\ref{tab:Parameters-for-Sarfati-Peleg-lik}
lists top eight candidates. The subsequent diagnostics are for the
Case No.1 (see Table.\ref{tab:Parameters-for-Sarfati-Peleg-lik}).

\subsubsection{Collective flash count and chorus}

The most fundamental characteristic of the collective SP-like flashing
is the emergence of periodic chorus flashes involving a large fraction
of the firefly population.

\begin{figure}
\begin{centering}
\includegraphics[width=0.5\textwidth]{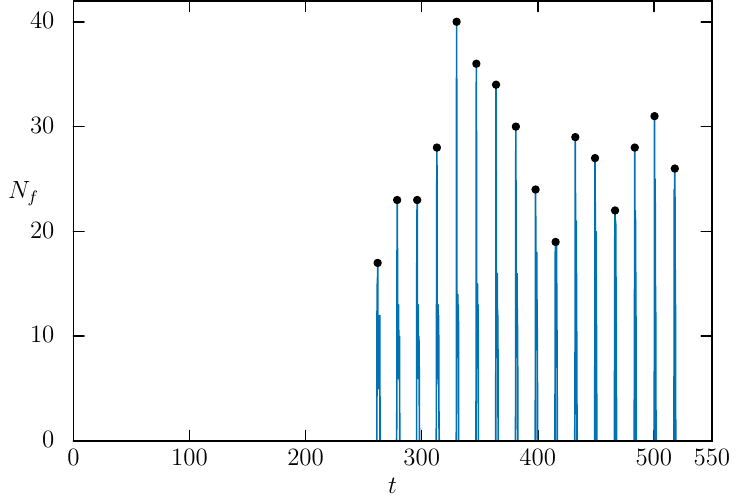}
\par\end{centering}
\caption{Collective flash and detected chorus. The blank part in the initial
phase is the transient phase during which self-organisation takes
place.}\label{fig:Collective-flash-and}
\end{figure}

For every oscillator, a \emph{flash} is defined as the instant at
which the state variable $y_{i}(t)$ spikes up or mathematically it
crosses a threshold value $y_{\textrm{flash}}$ from below. The threshold
$y_{\textrm{flash}}$ is chosen as a high quantile of the limit-cycle
trajectory of the uncoupled oscillator, so that only the sharp relaxation
spike is identified as a flash. The individual flash times are then
binned into a short time interval of duration $\Delta t_{\textrm{bin}}\ll T_{0}$.
If $N_{f}(t_{k})$ denotes the number of oscillators whose flash times
fall inside the interval $[t_{k},t_{k}+\Delta t_{\textrm{bin}})$,
then we have 
\begin{equation}
N_{f}(t_{k})=\sum^{N}_{i=1}\chi_{i}(t_{k}),\quad\chi_{i}(t_{k})=\left\{ \begin{array}{rl}
1, & t_{i}\in[t_{k},t_{k}+\Delta t_{\textrm{bin}}),\\
0, & \textrm{otherwise}.
\end{array}\right.
\end{equation}
So, $N_{f}(t)$ is simply the instantaneous number of fireflies flashing
simultaneously. This is the solid curve plotted in Fig.\ref{fig:Collective-flash-and}.

Now, for the chorus detection, we consider that the collective flash
count generally exhibits a sequence of pronounced maxima separated
by approximately one natural flashing period. So, to be a member of
a chorus, we must have 
\begin{equation}
\begin{array}{rcl}
N_{f}(t_{k}) & > & N_{f}(t_{k-1}),\\
 & \geq & N_{f}(t_{k+1}),
\end{array}
\end{equation}
so that the event corresponds to a genuine local peak. For a peak
to be classified as a chorus, the peak must also involve a sufficiently
large fraction of the population. So, if $f_{\textrm{chorus}}$ is
the fraction of the oscillators out of a total of $N$ oscillators,
which is the part of a chorus, then 
\begin{equation}
N_{f}(t_{k})\geq f_{\textrm{chorus}}N.
\end{equation}

These detected events are represented in Fig.\ref{fig:Collective-flash-and}
as black dots, which demonstrates that the heterogeneous delayed coupling
produces a sequence of well-defined collective flashing events that
persist throughout the observation interval. In these simulations,
we have fixed $f_{\textrm{chorus}}=0.65$. Here, this value is the
preferred chorus-participation fraction used during parameter selection
and is not the instantaneous peak-detection threshold. The chorus
peaks shown in Fig.\ref{fig:Collective-flash-and} were detected using
a minimum peak fraction of $f_{\textrm{peak}}=0.12$, corresponding
to $N_{f}\geq12$ for $N=100$. The pronounced periodic peaks indicate
that a substantial fraction of the oscillator population repeatedly
flashes almost simultaneously, while the detected chorus events (black
markers) provide a unique temporal reference for identifying the behaviour
of the remaining oscillators. Unlike a fully synchronized state, the
peak heights remain below the total population size, demonstrating
that a finite fraction of oscillators systematically flashes outside
the chorus. These non-participating oscillators subsequently constitute
the character population, whose temporal organization is analysed
in the subsequent diagnostics. Consequently, this figure establishes
the existence of a recurring chorus, which is the defining characteristic
of the collective flashing in a SP event and the basis for all subsequent
quantitative analyses \cite{SarfatiPeleg2022,Sarfati2021,Sarfati2023}.

\begin{table}
\caption{Simulation parameters\vr}\label{tab:Simulation-parameters}

\centering{}%
\begin{tabular}{|c|l|c|}
\hline 
\textbf{Parameter}\vrr & \multicolumn{1}{c|}{\textbf{Name}} & \textbf{Value}\tabularnewline
\hline 
\hline 
$\Delta_{c}$\vrr & chorus window & $0.18$\tabularnewline
\hline 
$d^{\textrm{min}}_{\textrm{char}}$\vrr & minimum character delay & $0.24$\tabularnewline
\hline 
$N_{{\cal R}}$\vrr & number of persistence runs & $5$\tabularnewline
\hline 
$\sigma_{\textrm{max}}$\vrr & max. allowed $\sigma_{i}$ during persistence runs & $0.22$\tabularnewline
\hline 
$\Delta t_{\textrm{bin}}$\vrr & chorus bin & $0.25$\tabularnewline
\hline 
\end{tabular}
\end{table}

\subsubsection{Cycle-resolved delay map}

Once the chorus events have been identified, the next objective is
to determine how every individual oscillator behaves relative to each
chorus. The defining characteristic of the SP event is that not all
fireflies flash simultaneously with the chorus. Instead, many individuals
repeatedly flash either before or after the collective event, thereby
exhibiting persistent temporal identities. The cycle-resolved delay
map provides a quantitative representation of this behaviour. Unlike
the collective flash count, which characterizes the behaviour of the
population as a whole, the cycle-resolved delay map follows the temporal
evolution of every individual oscillator over successive chorus cycles.

\begin{figure}
\begin{centering}
\includegraphics[width=0.5\textwidth]{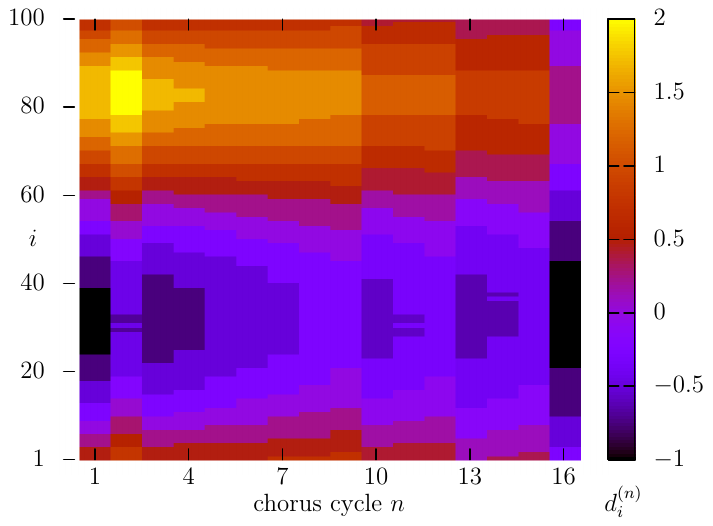}\hfill{}\includegraphics[width=0.5\textwidth]{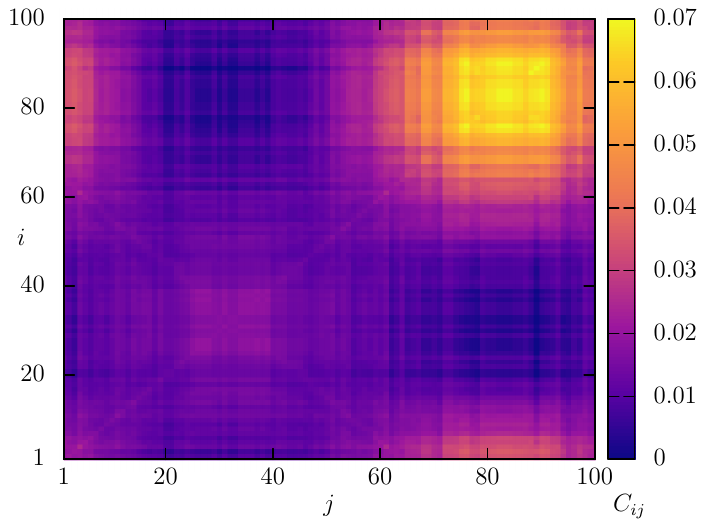}
\par\end{centering}
\caption{The delay map (left panel) and the covariance matrix map of the delay
map (right panel).}\label{fig:The-delay-map}
\end{figure}

Let us assume that the detected chorus times are given by 
\begin{equation}
t^{(1)}_{c},t^{(2)}_{c},\dots,t^{(N_{\textrm{chorus}})}_{c}
\end{equation}
and the successive flash times for the $i$th oscillator are 
\begin{equation}
t^{(1)}_{i},t^{(2)}_{i},\dots.
\end{equation}
Then, for each chorus event, every oscillator contributes exactly
one representative flash, namely the flash temporally closest to that
chorus event. Thus, the quantity 
\begin{equation}
t^{(n)}_{i}=\min_{t^{(m)}_{i}}\left|t^{(m)}_{i}-t^{(n)}_{c}\right|
\end{equation}
selects the flash from $i$th oscillator to the nearest chorus event.
So, we define the quantity 
\begin{equation}
d^{(n)}_{i}=t^{(n)}_{i}-t^{(n)}_{c}\label{eq:delay}
\end{equation}
as `character delay', which is the temporal delay of the $i$th oscillator
during $n$th chorus cycle. A value of $d^{(n)}_{i}=0$ indicates
that the oscillator participates in a chorus event, while $d^{(n)}_{i}\lessgtr0$
indicates that the character lags or precedes the chorus. The most
important feature, the delay map in Fig.\ref{fig:The-delay-map} (left
panel) is the persistence of the colour bands along the horizontal
direction. Such persistence indicates that the behavioural role is
not randomly reassigned from cycle to cycle. Instead, the oscillator
possesses a stable temporal identity. Conversely, frequent colour
changes along a row indicate switching between chorus and character
behaviour. Fig.\ref{fig:The-delay-map} (left panel) provides some
of the clearest numerical evidence for the emergence of persistent
character fireflies and constitutes the fundamental diagnostic of
the SP effect. The experiment of Sarfati and Peleg revealed that,
although a dominant chorus repeatedly emerged, numerous fireflies
consistently flashed slightly before or after the chorus, maintaining
recognizable temporal identities over many flashing cycles. The delay
map reproduces this behaviour quantitatively. Unlike the experimental
observations, which were primarily qualitative, the present analysis
assigns a numerical delay to every oscillator during every chorus
cycle. This allows the persistence, variability, covariance, and spatial
correlation of character behaviour to be investigated quantitatively. 

The right panel of Fig.\ref{fig:The-delay-map} statistically establishes
this delay effect and proves that the delay structure is collective
rather than individual. It calculates the second-order statistical
covariance matrix and displays in a heat-map. In order to calculate
the covariance matrix, we first compute the mean delay 
\begin{equation}
\bar{d}_{i}=\frac{1}{N_{\textrm{chorus}}}\sum^{N_{\textrm{chorus}}}_{n=1}d^{(n)}_{i}
\end{equation}
and define the delay fluctuations 
\begin{equation}
\delta d^{(n)}_{i}=d^{(n)}_{i}-\bar{d}_{i}.
\end{equation}
The covariance matrix is then defined as 
\begin{equation}
C_{ij}=\frac{1}{N_{\textrm{chorus}}}\sum^{N_{\textrm{chorus}}}_{n=1}\delta d^{(n)}_{i}\,\delta d^{(n)}_{j}.
\end{equation}
The calculation of covariance matrix is important because the SP phenomenon
is fundamentally a collective behaviour. If every character firefly
behaved independently, there would be no meaningful correlation structure.
The covariance matrix reveals that groups of oscillators share similar
delay dynamics, indicating that the temporal organization extends
beyond individual behaviour. Unlike the delay map, which describes
the behaviour of each oscillator individually, the covariance map
quantifies the relationship between every pair of oscillators. The
bright blocks correspond to groups of oscillators whose delay fluctuations
evolve cooperatively, whereas darker regions indicate weak or negative
correlations. As per the Sarfati-Peleg experiment, certain fireflies
repeatedly behave as characters rather than chorus participants. However,
the experiment did not quantify whether these character fireflies
fluctuate independently or exhibit collective temporal organization.
A structured covariance matrix indicates that the temporal behaviour
of character fireflies is not random but correlated across multiple
oscillators. In other words, persistent character behaviour is accompanied
by cooperative delay fluctuations, implying that the chorus-character
organization is a collective property of the interaction network rather
than the sum of independent individual responses.

\begin{figure}
\begin{centering}
\includegraphics[width=0.5\textwidth]{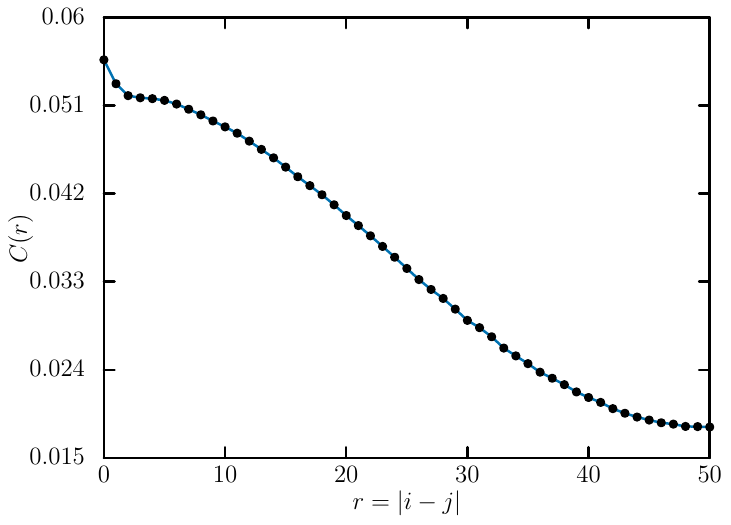}
\par\end{centering}
\caption{Spatial covariance function $C(r)$.}\label{fig:Spatial-covariance-function}
\end{figure}

As a further confirmation to the results from the delay map, we calculate
the `spatial covariance function' $C(r)$, which measures how similarly
two oscillators fluctuate as a function of their separation along
the ring. Large positive values indicate that oscillators separated
by the distance $r$ tend to fluctuate together. Accordingly, small
values indicate weak statistical dependence. It is computed as 
\begin{equation}
C(r)=\frac{1}{N_{r}}\sum_{r_{ij}=r}C_{ij},
\end{equation}
which is an average of all elements of covariance matrix $C_{ij}$
satisfying the condition $r_{ij}=r$ for a separation distance $r$.
In the above expression $N_{r}$ is the number of oscillator pairs,
separated by a distance $r$. The quantity $C(r)$ versus $r$ is
shown in Fig.\ref{fig:Spatial-covariance-function}, which shows a
monotonically decreasing behaviour. This shows that nearby oscillators
tend to oscillate in unison whereas oscillators separated by larger
distances fluctuate progressively more independently. The delay field
is therefore neither completely local nor completely global, instead
neighbouring oscillators form dynamically correlated groups whose
influence gradually weakens with distance. We note that the Sarfati-Peleg
experiment did not provide this information, though this should be
a natural consequence of the results.

\subsubsection{Role raster and character probability}

As a final Tier-I diagnostics, we present a raster plot of the role
of each oscillator and the switching probability between an oscillator
being the part of a chorus or being an individual oscillator -- a
`character'.

\begin{figure}
\begin{centering}
\includegraphics[width=0.5\textwidth]{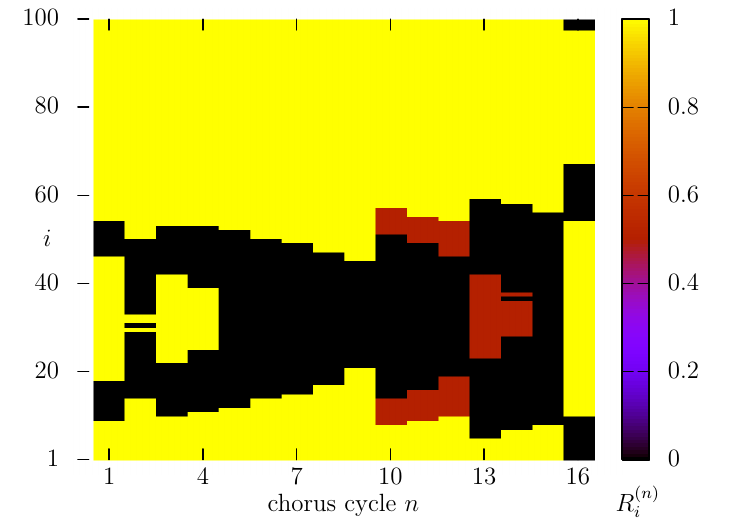}\hfill{}\includegraphics[width=0.5\textwidth]{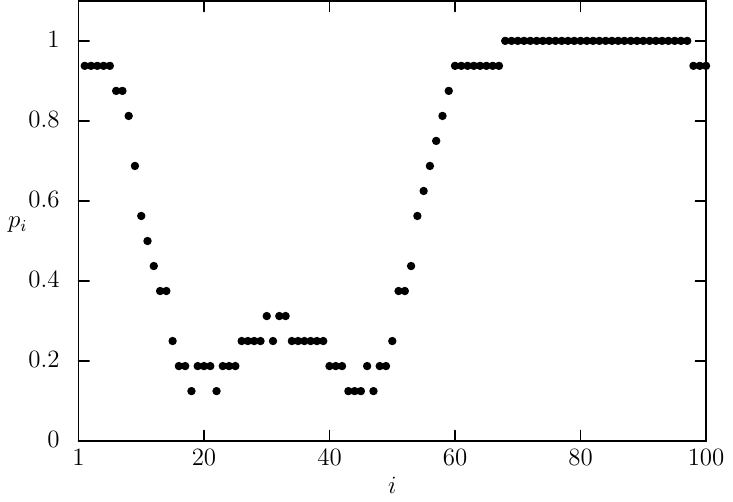}
\par\end{centering}
\caption{The raster plot of the role of the oscillators (left panel) and the
role-switching probability.}\label{fig:The-raster-plot}
\end{figure}

The raster plot represents the temporal evolution of behavioural states
of the oscillators. Note Eq.(\ref{eq:delay}), which defines $d^{(n)}_{i}$,
the delay for each oscillator with reference to the $n$th chorus
cycle. We now specifically define a delay threshold $d_{\textrm{th}}$
such that if 
\begin{equation}
\left|d^{(n)}_{i}\right|<d_{\textrm{th}},
\end{equation}
the oscillator is regarded as a member of the chorus and we assign
the \emph{behavioural state} of the oscillator $R^{(n)}_{i}$ as `chorus'
, otherwise the state is a `character' 
\begin{equation}
\begin{array}{rclll}
\left|d^{(n)}_{i}\right| & < & d_{\textrm{th}} & \Rightarrow & R^{(n)}_{i}=\textrm{chorus}\,:\chi^{(n)}_{i}=0,\\
 & \geq & d_{\textrm{th}} & \Rightarrow & R^{(n)}_{i}=\textrm{character}\,:\chi^{(n)}_{i}=1.
\end{array}
\end{equation}
The resultant raster plot is shown Fig.\ref{fig:The-raster-plot}
(left panel), which shows very distinctively that most of the members
of a `chorus' group remain in the group over many successive chorus
cycles, demonstrating that many oscillators preserve their behavioural
identity. Only a relatively small number of oscillators undergo transitions
between chorus and character states.

To supplement this observation, we also plot the `role-switching'
probability $p_{i}$ in the right panel of Fig.\ref{fig:The-raster-plot},
which measures the fraction of chorus cycles during which an oscillator
behaves as a `character' rather than as a `chorus' participant. The
probability $p_{i}$ is defined as 
\begin{equation}
p_{i}=\frac{1}{N_{\textrm{chorus}}}\sum^{N_{\textrm{chorus}}}_{n=1}\chi^{(n)}_{i}.
\end{equation}
The plot shows a cluster of oscillators are grouped with $p_{i}\sim0$,
indicating the fact that certain \emph{permanent }members of the chorus
never switch roles. This also true for a cluster of other group with
$p_{i}\approx1$, which shows that some oscillators always behave
as `chacarcters'. But there are two intermediary groups which keep
on switching roles between being a `character' or being a member of
a `chorus' and this behaviour is persistent over several cycles of
chorus. One of the principal conclusions of the Sarfati-Peleg experiment
was that individual fireflies preserve recognizable behavioural identities
over many flashing cycles. Some individual fireflies consistently
participate in the chorus, whereas others repeatedly behave as character.;
This numerical diagnostic provides evidence for this behavioural persistence
and also provides a quantitative measure..

\section{Analytical validation}

In this section, we try to identify the parameters which can be validated
by a reduced analytical model. Toward this, we define certain additional
parameters based on our above numerical results, which we shall call
our Tier-II diagnostics.

\subsection{Representative delay amplitude}

In this context, we define two parameters $\Delta_{c}$ and $d^{\textrm{min}}_{\textrm{char}}$
as the chorus window and minimum character delay, which define the
limits for $d^{(n)}_{i}$. In other words,
\begin{equation}
\begin{array}{rcllll}
\left|d^{(n)}_{i}\right| & \leq & \Delta_{c} &  &  & \Rightarrow\,\textrm{chorus}\\
\Delta_{c} & < & \left|d^{(n)}_{i}\right| & \leq & d^{\textrm{min}}_{\textrm{char}} & \Rightarrow\,\textrm{borderline}\\
\left|d^{(n)}_{i}\right| & > & d^{\textrm{min}}_{\textrm{char}} &  &  & \Rightarrow\,\textrm{character}
\end{array}
\end{equation}
The values of $\Delta_{c}$ and $d^{\textrm{min}}_{\textrm{char}}$
and other related parameters are given in Table.\ref{tab:Simulation-parameters}. 

We also define a \emph{persistence condition}, which determines whether
an oscillator maintains a stable, nonzero delay relative to the chorus
for several consecutive chorus cycles. Such an oscillator is treated
as a persistent character or non-switching character. So, for an oscillator
$i$, a simulation run ${\cal R}_{i}$ is accepted as persistent-character
run when it satisfies the following conditions
\begin{eqnarray}
\left|d^{(n)}_{i}\right| & > & d_{\textrm{min}}=\max\left(\Delta_{c},d^{\textrm{min}}_{\textrm{char}}\right),\\
{\cal R}_{i} & \geq & N_{{\cal R}},\\
\mu^{i}_{d} & \geq & \max\left(\Delta_{c},d^{\textrm{min}}_{\textrm{char}}\right),\\
\sigma_{i} & \leq & \sigma_{\textrm{max}},
\end{eqnarray}
where $N_{{\cal R}}$ is the number of minimum such runs, $\mu^{i}_{d}$
is the mean delay during such runs
\begin{equation}
\mu^{i}_{d}=\frac{1}{{\cal R}_{i}}\sum_{n\in{\cal R}_{i}}d^{(n)}_{i},
\end{equation}
and $\sigma_{i,\textrm{max}}$ are the cycle-to-cycle standard deviation
and its maximum permissible value
\begin{equation}
\sigma_{i}=\sqrt{\frac{1}{{\cal R}_{i}-1}\sum_{n\in{\cal R}_{i}}\left(d^{(n)}_{i}-\mu_{i}\right)^{2}}.
\end{equation}

If $\mathcal{S}$ is the set of oscillators satisfying the persistence
condition, we define representative delay amplitude as
\begin{equation}
d_{\mathrm{char,num}}=\underset{i\in\mathcal{S}}{\operatorname{median}}\left|\mu^{i}_{d}\right|.\label{eq:d-char-num}
\end{equation}
For the numerical case considered here, this auxiliary post-processing
statistics gives 
\begin{equation}
d_{\mathrm{char,num}}=0.501667.\label{eq:d-char-num-value}
\end{equation}
This value is obtained entirely from the numerical delay tracks generated
by the original oscillator model. The complete extraction procedure
is outlined in Appendix A.

\subsection{Controlled settle-kick-observe validation }

As the title suggests, we perform a controlled settle-kick-observe
experiment on a quasi-stationary numerical SP-like state. The numerical
network is first allowed to evolve without an externally imposed perturbation
so that its initial transient decays and a collective flashing state
is established. The settling time is chosen as
\begin{equation}
t_{\textrm{settle}}=260
\end{equation}
in the numerical time units of the simulation.

At $t=t_{\mathrm{settle}}$, a controlled phase perturbation, hereafter
referred to as a kick, is applied to a contiguous block of oscillators.
Both positive and negative kicks are considered. Assume that $(x_{i},y_{i})$
be the state of oscillator $i$ immediately before the kick, and let
$(c_{x},c_{y})$ denote the centre of the numerically constructed
single-oscillator limit cycle. The instantaneous geometric phase is
defined as
\begin{equation}
\phi_{i}=\operatorname{atan2}\left(y_{i}-c_{y},\,x_{i}-c_{x}\right).
\end{equation}
For every oscillator in the perturbed block, the kick changes this
phase according to
\begin{equation}
\phi_{i}\longrightarrow\phi_{i}+s\Delta\phi,\qquad s=\pm1,
\end{equation}
where $s=+1$ denotes a positive kick and $s=-1$ denotes a negative
kick. The shifted phase is then mapped back onto the same numerically
constructed limit cycle
\begin{equation}
(x_{i},y_{i})\longrightarrow\left(x_{0}(\phi_{i}+s\Delta\phi),y_{0}(\phi_{i}+s\Delta\phi)\right),
\end{equation}
where $(x_{0},y_{0})$ denote a point on the limit cycle of an individual
oscillator. Therefore, the kick is a phase-resetting perturbation.
It does not place the oscillator on a larger or smaller orbit, nor
does it directly change the intrinsic oscillator parameters or its
natural period. Instead, it instantaneously relocates the oscillator
to a different point on the same limit cycle.

After the kick, the system is allowed to evolve under the original
delayed network equations. The oscillators outside the kicked block
are used as a reference population for reconstructing the chorus times.
The delay of the kicked block is then measured relative to this surviving
reference chorus. A negative block delay,
\begin{equation}
d<0,
\end{equation}
means that the kicked population flashes before the reference chorus,
whereas a positive delay,
\begin{equation}
d>0,
\end{equation}
means that it flashes after the reference chorus.

\section{A dynamical model}

We now examine whether the essential features of our numerical analogue
of the SP phenomenon can be captured by a reduced dynamical model.
As demonstrated in the previous section, we have successfully reproduced
SP-like behaviour in a ring of firefly-inspired relaxation oscillators
coupled through a distributed communication time delay. Although the
Sarfati-Peleg experiment does not explicitly identify communication
delay -- or even suggest that it is experimentally measurable --
our numerical results show that distributed communication delay is
an effective ingredient of our numerical model. Without it, our present
model and parameter search did not reproduce the observed chorus--character
behaviour. Before proceeding further, however, a few clarifications
are necessary.

In the Sarfati-Peleg experiment, the fireflies are not fixed in space;
they move freely, change their orientations, and continuously modify
their interaction neighbourhoods. Moreover, several environmental
factors, such as visual occlusion and the resulting changes in interaction
topology, evolve continuously in time. The communication delays between
individual fireflies are also unknown and, at present, are not experimentally
measurable. It is therefore unrealistic to expect a mathematical model
to incorporate every detail of the natural environment in which the
SP phenomenon is observed. Rather, the objective of a theoretical
model is to identify the minimal set of dynamical ingredients required
to reproduce the essential phenomenon. Since our numerical model with
distributed communication delay successfully captures the key qualitative
features of the SP observations, our aim in this section is to \emph{phenomenologically}
derive a reduced dynamical description that explains the underlying
mechanism responsible for this numerical analogue of the SP phenomenon.

\subsection{Tri-stability}

We note that there are two delays in our system, which \emph{must
not} be confused. The pair-wise distributed communication time delay
$\tau_{ij}$ is a quenched model parameter determined by the pair
separation and a random heterogeneous component, 
\begin{equation}
\tau_{ij}\equiv\tau_{ij}(s_{ij},\xi_{ij}),
\end{equation}
where $s_{ij}$ is the distance between the $i$th and $j$th oscillators.
The coupling strength and radius are also fixed in the model and do
not evolve dynamically. Then there is the quantity $d^{(n)}_{i}$
for each oscillator, which is the delay $i$th oscillator has relative
to the nearest chorus event, which varies with chorus cycle and oscillator
position, and so 
\begin{equation}
d^{(n)}_{i}\equiv d^{(n)}_{i}(\{\tau_{ij}\};s,t),
\end{equation}
where $\{\tau_{ij}\}$ denotes the set of pairwise communication delays
affecting oscillator $i$. We now take a continuum limit and introduce
the two continuous variables $(s,t)$ and define a `delay field' $d(s,t)$,
so that 
\begin{equation}
d_{i}\to d(s,t),
\end{equation}
and try to form an evolution equation for this delay field. Naturally
$d=0$ would mean that it is a chorus event. Nonzero equilibria of
$d\neq0$ would mean characters flashing not as a part of the chorus.
Essentially, our dynamical model should represent a stable equilibrium
point at $d=0$, which coexists with stable $d\neq0$ equilibrium
points. This suggests a ``subcritical pitchfork bifurcation with
saturation'' \cite{strogatz2015nonlinear,guckenheimer1983nonlinear,bizyaeva2023fast}
or a catastrophic or generalized pitchfork bifurcation, leading to
tri-stability.

To build our model around this hypothesis, we first assume a `snapping'
effect: a character oscillator slightly displaced temporally from
a chorus is pulled back by the collective synchronisation. So, by
itself, we have 
\begin{equation}
\dot{d}=-ad,\quad a>0,
\end{equation}
which has a stable solution at $d=0$. So, this equation represents
a global synchrony. However, we need a nonlinear opposing term as
we have seen that the group resists global synchrony for sufficiently
weak coupling. The simplest nonlinear term would be a cubic nonlinearity,
so that our equation becomes
\begin{equation}
\dot{d}=-ad+bd^{3},\quad b>0.
\end{equation}
This term represents an instability in the sense that so long as $d$
is small, the linear term dominates, but for larger $d$, the instability
would pull the system away from an equilibrium. But, we observe that
in the absence of a global synchrony $(a=0)$, the solution of the
above equation is explosive, which is \emph{not} physical and we need
to have a bounded $d$. This calls for a quintic nonlinear saturation
term, 
\begin{equation}
\dot{d}=-ad+bd^{3}-cd^{5},\quad c>0.
\end{equation}
The quintic term represents nonlinear saturation of the instability.
In summary, the local evolution part 
\begin{equation}
f(d)=-ad+bd^{3}-cd^{5},\quad(a,b,c)>0,
\end{equation}
is the standard normal form of a subcritical pitchfork bifurcation
with saturation \cite{Strogatz2018,CrossHohenberg1993}.

\subsection{Delay diffusion and nonlocal interaction}

We realise that communication delay $\tau$ makes each oscillator
respond to a neighbourhood average. If neighbouring oscillators have
different chorus-relative delays, the delayed input tends to reduce
those differences over repeated cycles. Therefore the delay field
$d(s,t)$ is smoothed by the coupling, which is like a diffusion ${\cal D}(s,t)$.
We can think of this just like a diffusion of heat energy in a conductor,
only in this case, the delay is itself diffusing to the neighbourhood.
The last effect that we would like to incorporate is the nonlocal
distributed-delay interaction term ${\cal L}[d]$. With these two
terms, we can write the final evolution equation for $d(s,t)$ as
\begin{equation}
\dot{d}=-ad+bd^{3}-cd^{5}+{\cal D}(s,t)+{\cal L}[d],\label{eq:evol}
\end{equation}
where ${\cal D}(s,t)$ is the diffusion term 
\begin{eqnarray}
{\cal D}(s,t) & = & D_{d}\frac{\partial^{2}d}{\partial s^{2}},\\
{\cal L}[d] & = & \Lambda\,{\cal I}(s,t),
\end{eqnarray}
with $D_{d}$ as the diffusion coefficient. Assume that an oscillator
at $s$ receives information from every other oscillator at $s'$.
But as the communication delay $\tau$ is a part of the receiving
process, the contribution of the delay field for the oscillator at
$s$ must be of the form $d(s',t-\tau(s,s'))$. But, every oscillator
communicates with many other neighbours and every neighbour contributes
a delayed signal, so the total information received is a weighted
sum giving rise to 
\begin{equation}
{\cal I}(s,t)=\int_{\mathfrak{R}}K(r)\,[d(s+r,t-\tau(r))-d(s,t)]\,dr,
\end{equation}
where $r=s'-s$ and the kernel $K(r)$ represents how strongly the
oscillator at $s'$ influences the oscillator at $s$ or in other
words it contains information about spatial communication topology.
There can be several possible candidates for $K(r)$ including nearest
neighbour, global, finite-range, Gaussian, to name a few. In this
context, $\Lambda$ represents communication susceptibility or gain,
which controls how strongly $K(r)$ should be incorporated into the
model. We also assume the kernel to be normalised 
\begin{equation}
\int_{\mathfrak{R}}K(r)\,dr=1.
\end{equation}

\subsection{Memory effect and two-field evolution model}

However, as we have seen that certain character oscillators tend remain
away from the chorus, the diffusion of the delay field is somewhat
opposed by this `memory' effect, which can be represented by a memory
term $m(s,t)$, which helps each oscillator retain its past status
of being a character or a member of a chorus. But this `memory' effect
cannot also be completely static and must evolve slowly over time.
As this `memory' also affects the nonlinear growth and saturation,
the cubic and quintic terms $(+\,bd^{3}-cd^{5})$ of our present evolution
equation should be part of this `memory evolution'. To this effect,
we propose the following two-field analytical model for our numerical
analogue
\begin{eqnarray}
\frac{\partial}{\partial t}d(s,t) & = & -a\,d(s,t)+\chi\,m(s,t)+D_{d}\frac{\partial^{2}}{\partial s^{2}}d(s,t)\nonumber \\
 &  & +\,\Lambda\int K(r)\,[d(s+r,t-\tau(r))-d(s,t)\,dr,\label{eq:evol1}\\
\tau_{m}\frac{\partial}{\partial t}m(s,t) & = & -m(s,t)+G[d]+D_{m}\frac{\partial^{2}}{\partial s^{2}}m(s,t),\label{eq:evol2}
\end{eqnarray}
where $\chi$ controls the strength of the memory term, $\tau_{m}>0$
is the characteristic `memory decay' time, $G[d]$ represents a delay
to the memory response, and the last term in Eq.(\ref{eq:evol2})
is a memory diffusion term with $D_{m}$ as the `memory diffusion
coefficient'. The delay term in memory response $G[d]$ is designed
so that it becomes equivalent to the cubic nonlinear and quintic saturation
terms $(+bd^{3}-cd^{5})$ of Eq.(\ref{eq:evol}) when $\tau_{m}\to0$.
This term is thus given by
\begin{equation}
G[d]=\frac{d\left(1+\beta_{m}d^{2}\right)}{1+\gamma_{m}d^{4}},
\end{equation}
where $(\beta,\gamma)_{m}$ are constants. In the limit of $\tau_{m}\to0$,
from Eq.(\ref{eq:evol2}), we have approximately
\begin{equation}
m=G[d]\simeq d+\beta_{m}d^{3}-\gamma_{m}d^{5}+\cdots,
\end{equation}
where we have neglected the memory diffusion term and expanded $G[d]$
for small $d$. Substituting the above expression in Eq.(\ref{eq:evol1})
provides us the cubic nonlinear and the quintic saturation terms.
Naturally in these limits, Eqs.(\ref{eq:evol1},\ref{eq:evol2}) supports
the underlying properties of a subcritical pitchfork bifurcation.
The constants $(\beta,\gamma)_{m}$ can be thought to be the strengths
of the cubic nonlinearity and quintic saturation terms, respectively.

\subsection{Equilibrium and stability}

\subsubsection{Homogenous state}

We consider a spatially uniform, time-independent (steady state) delay
field as well as a constant memory field
\begin{equation}
d(s,t)=d_{\star},\quad m(s,t)=m_{\star}.
\end{equation}
Naturally all the diffusion term vanishes as well as the nonlocal
interaction term as ${\cal I}[d_{\star}]=0$. The equilibrium equations
become
\begin{equation}
\left.\begin{array}{rcl}
0 & = & -ad_{\star}+\chi m_{\star},\\
0 & = & -m_{\star}+G(d_{\star}),
\end{array}\right\} \Rightarrow ad_{\star}=\chi G(d_{\star}).
\end{equation}
Note that the non-zero branches represent the character solutions,
which provides the constraint on $a$,
\begin{equation}
a=\chi\left(\frac{1+\beta_{m}z}{1+\gamma_{m}z^{2}}\right),\quad d_{\star}\neq0,\label{eq:non-zero}
\end{equation}
where $z=d^{2}_{\star}>0$. Assuming the Sarfati-Peleg state to be
quasi-stationary state of the ring of oscillators to which they have
settled down, a good estimate for $d_{\star}$ would be to use
\begin{equation}
d_{\star}\equiv d_{\mathrm{char,num}}\simeq0.502\Rightarrow z\equiv z_{\star}\simeq0.252,
\end{equation}
the representative delay amplitude already calculated from the numerical
model.
\begin{table}
\caption{Equilibrium points $(d_{\star})$ and their stability for the homogeneous
state of the two-field model.\vr}\label{tab:Equilibrium-points-}

\centering{}%
\begin{tabular}{|c|l|}
\hline 
$\bm{d_{\star}}$\vrr & \multicolumn{1}{c|}{\textbf{Stability and Role}}\tabularnewline
\hline 
\hline 
$-0.502$\vrr & Stable, Negative delayed character state\tabularnewline
\hline 
$-0.445$\vrr & Unstable, Intermediate negative delayed character state\tabularnewline
\hline 
$0$\vrr & Stable, Chorus state\tabularnewline
\hline 
$+0.502$\vrr & Stable, Postive delayed character state\tabularnewline
\hline 
$+0.445$\vrr & Unstable, Intermediate positive delayed character state\tabularnewline
\hline 
\end{tabular}
\end{table}

Eq.(\ref{eq:non-zero}) can be re-arranged as
\begin{equation}
a\gamma_{m}z^{2}-\chi\beta_{m}z+(a-\chi)=0.
\end{equation}
Using $z=z_{\star}$ in the constraint for $a$ in Eq.(\ref{eq:non-zero}),
it can be easily shown that $z=z_{\star}$ is a root of the above
equation. The other root $\tilde{z}_{\star}$ can be shown to be equal
to
\begin{equation}
\tilde{z}_{\star}=\frac{\beta-\gamma z_{\star}}{\gamma(1+\beta z_{\star})}.
\end{equation}
Thus, we have five homogeneous equilibrium points
\begin{equation}
d_{\star}=0,\quad d_{\star}=\pm\sqrt{z_{\star}},\quad d_{\star}=\pm\sqrt{\tilde{z}_{\star}}.
\end{equation}

\subsubsection{Stability}

We perform a standard linear stability analysis, for which the Jacobain
at the equilibrium point $d=d_{\star}$ is given by
\begin{equation}
J=\left(\begin{array}{cc}
-a & \chi\\
G'(d_{\star})/\tau_{m} & -1/\tau_{m}
\end{array}\right),
\end{equation}
for which the trace and determinant are given by
\begin{eqnarray}
\textrm{tr}(J) & = & -a-\frac{1}{\tau_{m}},\\
\det(J) & = & \frac{a-\chi G'(d_{\star})}{\tau_{m}},
\end{eqnarray}
which determine the linear stability of the equilibrium points. As
$G'(d_{\star})=1$ for $d_{\star}=0$, the chorus equilibrium stable
only if $a>\chi$, as both $(a,\tau_{m})>0$. It can be further shown
that for $a>\chi$, the larger magnitude pair of roots between $\pm\sqrt{z_{\star}},\pm\sqrt{\tilde{z}_{\star}}$
are always stable and the lower magnitude pair is always unstable.
Thus we arrive at the tri-stability of the homogenous state of the
two-field model for $a>\chi$.

We note that the values of $\beta_{m},\gamma_{m}$, and $\tau_{m}$
are selected by comparing the two-field model with the numerical delay
tracks, from which we fix these values at $\beta_{m}=1$, $\gamma_{m}=2$,
and $\tau_{m}=17$. For details about the process of selection, please
see Appendix B. For the best fitted values of coefficients of the
reduced model, the equilibrium values with their stability are given
in Table.\ref{tab:Equilibrium-points-}.

\begin{figure}
\begin{centering}
\includegraphics[width=0.5\textwidth]{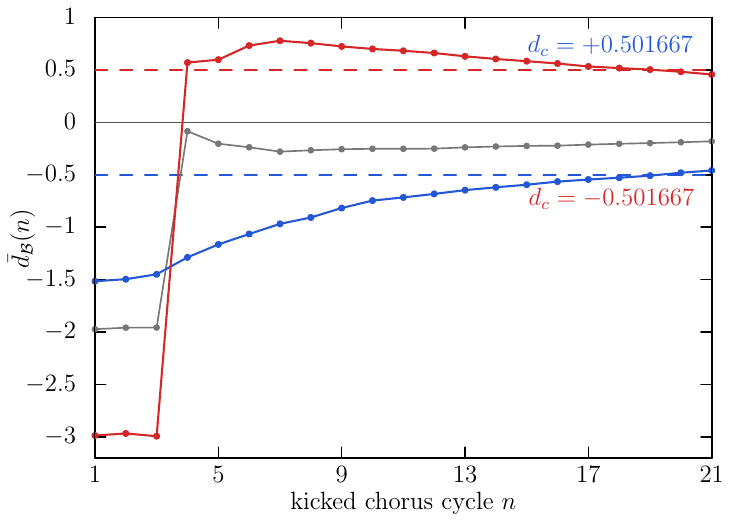}
\par\end{centering}
\caption{Settle-kick evolution of the block-median delay. The cycle-resolved
median delay $\bar{d}_{\mathcal{B}}(n)$ of a spatial block is measured
relative to chorus times reconstructed from the complementary unperturbed
population. Gray, blue, and red curves show the zero-kick control,
negative-kick trajectory $\Delta\phi=-0.008$, and positive-kick trajectory
$\Delta\phi=+0.00575$, respectively. Dashed horizontal lines denote
the calibrated analytical character levels $d_{c}=\pm0.501667$. Opposite
kicks produces opposite signed delay sectors, and both trajectories
pass close to the corresponding reference level during the finite
observation interval.}\label{fig:Settle-kick-evolution-of}
\end{figure}

\subsection{Validation}

In Fig.\ref{fig:Settle-kick-evolution-of}, we show the results of
the controlled settle-kick-observe procedure on the quasi-stationary
numerical SP state, which shows that a that a spatially kicked block
develops a signed delay relative to the unperturbed reference population..
The numerical network is first allowed to evolve until 
\begin{equation}
t_{\mathrm{settle}}=260,
\end{equation}
after which a phase kick is applied to a contiguous spatial block
$\mathcal{B}$ of oscillators. The oscillators outside the kicked
block constitute the reference population and are used to reconstruct
the post-kick chorus times. For every post-kick chorus cycle $n$,
we define the signed block-median delay 
\begin{equation}
\bar{d}_{\mathcal{B}}(n)=\textrm{median}_{i\in{\cal B}}d^{(n)}_{i},
\end{equation}
where 
\begin{equation}
d^{(n)}_{i}=t^{(n)}_{i}-t^{(n)}_{c}
\end{equation}
is the delay of oscillator $i$'s assigned flash, relative to the
chorus time determined from the unperturbed reference population.
The quantity $\bar{d}_{\mathcal{B}}(n)$ therefore measures whether
the kicked block flashes earlier or later than the surviving reference
chorus. Under the sign convention used here, 
\begin{equation}
\begin{array}{rclll}
\bar{d}_{\mathcal{B}}(n)>0 & \Longrightarrow & t^{(n)}_{i}>t^{(n)}_{c} & \Longrightarrow & \text{the kicked block lags the chorus},\\[4pt]
\bar{d}_{\mathcal{B}}(n)<0 & \Longrightarrow & t^{(n)}_{i}<t^{(n)}_{c} & \Longrightarrow & \text{the kicked block precedes the chorus}.
\end{array}
\end{equation}
The spontaneously self-organized, un-kicked numerical case discussed
earlier, contains a reference chorus and persistent positive-delay
character populations. For that case, we have 
\begin{equation}
0.315\lesssim\mu_{i}\lesssim0.835,
\end{equation}
and their representative absolute median is 
\begin{equation}
d_{\mathrm{char,num}}=\textrm{median}_{i\in{\cal S}}|\mu_{i}|\simeq0.502,
\end{equation}
where ${\cal S}$ is the set of oscillators whose numerical delay
tracks satisfy the persistent-character conditions. In our numerical
analogue of the SP phenomenon, no persistent negative-delay population
is identified in the un-kicked state. However, this does not, by itself,
contradict the reflection-symmetric analytical model as a particular
numerical trajectory need not simultaneously occupy every stable branch
permitted by the reduced equations. Nevertheless, it means that the
negative analytical branch is not directly demonstrated by the numerical
experiment.

Fig.\ref{fig:Settle-kick-evolution-of} shows how controlled positive
and negative phase kicks to a spatial block generate, respectively,
delayed and advanced block trajectories relative to a surviving reference
chorus. During the finite observation interval, these trajectories
pass close to the calibrated character-delay levels $d_{c}=\pm0.502$.
The result demonstrates controlled accessibility of both signed delay
sectors in the microscopic numerical dynamics. It is also important
to note that the original SP phenomenon demonstrates a dominant chorus
coexisting with persistent off-phase character flashes, but it does
not classify those character populations as two symmetric branches
with fixed delays $+d_{c}$ and $-d_{c}$. The experimental delay
is measured relative to a nearby chorus beat, and the sign may depend
on whether the preceding or following chorus is chosen as the reference.
Consequently, the experimental observations support persistent nonzero
character delays but do not independently establish a symmetric pair
of positive- and negative-delay equilibria. The settle-kick-observe
experiment provides an auxiliary test of whether both signed delay
sectors can be accessed within the our numerical dynamics. The controlled
kicks of opposite signs produce advanced and delayed block trajectories
that is quite close to the two sign-related character levels indicated
by the reduced model. In particular, the negative kick results a negative-delay
sector. Within the present simulation time, this result can be assumed
as a proof of the tri-stable state, but should be interpreted as finite-time
consistency only, rather than as a proof of asymptotic tri-stability.
The cycle-resolved trajectories continue to drift slowly after approaching
$d=\pm d_{\mathrm{char,num}}$.

\section{Epilogue -- summary and conclusions}

In this work, we have investigated a network of identical, nonlocally
coupled firefly oscillators that exhibits two distinct chimera-like
regimes depending on the structure of the communication delay. This
work is inspired by a recently observed chimera-like phenomenon in
a natural swarm of \emph{Photuris frontalis} fireflies \cite{SarfatiPeleg2022}.
To mimic this natural firefly swarm, we have modeled each firefly
using a numerical relaxation oscillator based on a chemical trigger
\cite{SaikiaBora2020}. It is important to realize that this numerical
firefly model can successfully generate experimentally observed firefly
flashes across different ambient temperatures.

In the uniform-delay case, the network shows the coexistence of frequency-locked
and frequency-drifting populations. The global coherence and the width
of the incoherent domain are found to vary periodically, demonstrating
that the observed state is a breathing chimera rather than a stationary
chimera. The characteristic breathing frequency is comparable to the
mean flashing frequency of the oscillators. This suggests that the
periodic expansion and contraction of the incoherent domain are linked
to the underlying relaxation-oscillation cycle.

In the case of heterogeneous communication delays, the same network
produces a qualitatively different organizational state akin to the
SP-like state. This state shows the coexistence of a dominant, periodically
recurring chorus with character oscillators that consistently flash
outside the chorus window. This SP-like state is characterized mainly
by persistent relative flash timing rather than by a separation of
the oscillators according to their long-time mean frequencies. Thus,
it is closer to a phase- or timing-organized chimera than to the frequency-drifting
chimera obtained with uniform delay. Cycle-resolved delay maps, role
rasters, character probabilities, and spatial covariance measurements
show that some oscillators remain persistent chorus members, some
maintain character roles, and others switch between these roles over
successive chorus cycles, just as in the natural SP phenomenon. The
persistent character oscillators have nonzero delays relative to the
chorus. From the numerical delay tracks, their representative delay
magnitude is estimated as $d_{\mathrm{char,num}}\approx0.502$. The
numerical results indicate that distributed communication delay is
an effective mechanism for producing persistent temporal identities
among identical oscillators. In the present model and parameter search,
the SP-like chorus-character organization was obtained only with this
delay heterogeneity.

Finally, we have proposed a reduced analytical model to examine whether
the SP-like state can be understood in terms of fundamental nonlinear
dynamical principles. The reduced analytical model was constructed
using a chorus-relative delay field and a memory field. Its local
dynamics is based on a saturated subcritical pitchfork structure that
possesses five homogeneous equilibria. Of these equilibria, the chorus
state $d_{\star}=0$ and the two character states $d_{\star}\approx\pm0.502$
are stable, whereas the intermediate states $d_{\star}\approx\pm0.445$
are unstable. The reduced model therefore exhibits a tri-stable organization
consisting of one chorus branch and two sign-related character branches.
To examine whether the two character-delay sectors are accessible
in the original oscillator network, we performed a controlled settle--kick--observe
experiment. The network was first allowed to settle into a quasi-stationary
collective flashing state, after which a positive or negative phase
kick was applied to a contiguous block of oscillators. Following the
kick, the unperturbed oscillators were used to reconstruct the reference
chorus, and the median delay of the kicked block was measured relative
to it. Positive and negative kicks generated delayed and advanced
block trajectories, respectively, which passed close to the analytical
character levels $d=\pm0.502$. The settle--kick--observe results
demonstrate the finite-time accessibility of both signed delay sectors
in the microscopic numerical dynamics. However, because the trajectories
continue to drift slowly, this should be interpreted only as finite-time
consistency with the proposed tri-stable structure rather than as
proof of asymptotic tri-stability. We also note that the experimental
results of Sarfati and Peleg’s observation of a chorus--character
chimera in natural fireflies show only a bistable structure rather
than a tri-stable one. However, as we understand it, the analytical
model is highly reduced and is proposed to explain the basic nonlinear
dynamical structure. Accordingly, we do not expect the experimental
results to fully reproduce the analytical results. In this context,
we argue that, experimentally, the firefly network is able to access
only two of the three branches -- one character branch and one chorus
branch -- rather than the analytically established tri-stable state.

Overall, our work shows that the same identical-oscillator framework
can reproduce both a breathing phase--frequency chimera and an SP-like
chorus--character state. The latter provides a numerical mechanism
for the persistent temporal identities observed in natural synchronous
firefly swarms, as reported by Sarfati and Peleg \cite{SarfatiPeleg2022}.

\bmhead{Acknowledgements}

This work is carried out under the research grant RAC-S/GU/2024/4/74 of ISRO Regional Academic Centre for Space, Gauhati University.

\section*{Declarations}

\begin{itemize}
\item {\bf Funding} This work is supported by ISRO's RESPOND programme under the research grant RAC-S/GU/2024/4/74 of ISRO Regional Academic Centre for Space, Gauhati University.

\item {\bf Competing interests} The authors declare no competing interests.
\item {\bf Ethics approval} Not applicable
\item {\bf Consent for publication} Not applicable
\item {\bf Data availability} Not applicable
\item {\bf Materials availability} Not applicable
\item {\bf Code availability} https://doi.org/10.5281/zenodo.24150199
\item {\bf Author contribution} DS conceptualised the idea and did the initial research. MPB worked out detailed methodology and drafted the manuscript. DD carried out the simulations and drawn the figures. KB supervised the work. The final version was approved all the authors.

\end{itemize}

\begin{appendices}

\section{Persistent-character selection}

A persistence run is accepted when 
\begin{equation}
{\cal R}_{i}\geq N_{{\cal R}},\qquad|\mu^{i}_{d}|\geq\max\left(\Delta_{c},d^{\textrm{min}}_{\textrm{char}}\right),\qquad\sigma_{i}\leq\sigma_{\textrm{max}}.
\end{equation}
If more than one run is admissible, the longest is selected and equal-length
runs are resolved in favour of the smaller standard deviation.

For this numerical realization, the selected oscillator set is
\begin{eqnarray}
\mathcal{S} & = & \{1,2,3,4,5,6,7,8,9,10,11,12,53,54,55,56,57,58,\nonumber \\
 &  & 59,60,61,62,63,64,65,66,67,98,99,100\},
\end{eqnarray}
and hence $|\mathcal{S}|=30$. Their accepted run lengths are 
\begin{eqnarray}
\{15,15,15,15,15,14,14,12,11,8,7,5,5,8,9,11,11,12,14,15,\\
15,15,15,15,15,15,15,15,15,15\}.\nonumber 
\end{eqnarray}
The associated standard deviations lie between $0.0793$ and $0.1668$,
within the imposed upper limit of $\sigma_{\textrm{max}}$.

To obtain a branch-independent amplitude $a_{(j)}$, their absolute
values are used. As there are thirty tracks, the median is the average
of the fifteenth and sixteenth ordered values,
\begin{equation}
a_{(15)}=a_{(16)}=0.502.
\end{equation}
The quantity $d_{\mathrm{char,num}}$ is thus given by
\begin{equation}
d_{\mathrm{char,num}}=\frac{a_{(15)}+a_{(16)}}{2}\simeq0.502.
\end{equation}

\section{Estimating the reduced coefficients}

\subsection{Methodology and constraints}

We use the following possible sets for $\beta_{m}$ and $\gamma_{m}$,
\begin{eqnarray}
\beta_{m} & \in & \{0.25,0.5,1,2,4,8\},\label{eq:set1}\\
\gamma_{m} & \in & \{0.01,0.05,0.1,0.25,0.5,1,2\},\label{eq:set2}
\end{eqnarray}
of which the elements are roughly double of its predecessor and covers
a range from being small to modestly large. The set for $\tau_{m}$
is chosen as
\begin{equation}
\tau_{m}=\alpha T_{c},\quad\alpha\in\{0.5,1,2,4\},\label{eq:set3}
\end{equation}
where $T_{c}$ is the median chorus period extracted from the numerical
simulation,
\begin{equation}
T_{c}=\textrm{median}\left(t^{(n+1)}_{c}-t^{(n)}_{c}\right).
\end{equation}
For each triplet $(\beta_{m},\gamma_{m},\tau_{m})$, the memory-response
function $G[d]$ is calculated by using the relation
\begin{equation}
G[d]=\frac{d\left(1+\beta_{m}d^{2}\right)}{1+\gamma_{m}d^{4}}
\end{equation}
using each $d=d^{(n)}_{i}$. The numerical memory is constructed from
the delay tracks using the following equation
\begin{equation}
m^{(n+1)}_{i}=e^{-\Delta t_{n}/\tau_{m}}m^{(n)}_{i}+\left(1-e^{-\Delta t_{n}/\tau_{m}}\right)G[d^{(n)}_{i}],\label{eq:mem}
\end{equation}
where
\begin{equation}
\Delta t_{n}=t^{(n+1)}_{c}-t^{(n)}_{c}.\label{eq:delat}
\end{equation}
Note that $(n)$ represents the $n$th chorus. Thus, every candidate
triplet generates a different reconstructed memory field. Eq.(\ref{eq:mem})
is derived from Eq.(\ref{eq:evol2}) as follows.

Between each chorus, we have the memory equation for the $i$th oscillator
as
\begin{equation}
\tau_{m}\frac{dm_{i}}{dt}=-m_{i}+G[d^{(n)}_{i}],
\end{equation}
where we have neglected the memory-diffusion term. Multiplying the
above equation by the integrating factor $e^{t/\tau_{m}}$, we have
\begin{equation}
\frac{d}{dt}\left[e^{t/\tau_{m}}m_{i}(t)\right]=\frac{G[d^{(n)}_{i}]}{\tau_{m}}e^{t/\tau_{m}}.
\end{equation}
Integrating the above expression from $t^{(n)}_{c}$ to $t^{(n+1)}_{c}$,
we obtain Eq.(\ref{eq:mem}). In Eq.(\ref{eq:mem}), $m^{(n)}_{i}=m_{i}(t^{(n)}_{c})$.
Now, for each $(\beta_{m},\gamma_{m})$, we define the quantity 
\begin{equation}
R_{c}=\frac{1+\beta_{m}d^{2}_{c}}{1+\gamma_{m}d^{4}_{c}},\quad d_{c}=d_{\textrm{char},\textrm{num}}.\label{eq:char-eqlb}
\end{equation}
The equilibrium constraint on $a$ is given by Eq.(\ref{eq:non-zero}),
\begin{equation}
a=\chi R_{c}.\label{eq:constraint}
\end{equation}
For a stable chorus, we \emph{must} have $a>\chi$, as indicated in
the stability analysis. So, every $(\beta_{m},\gamma_{m},\tau_{m})$
triplet produces its own fitted reduced model

To estimate other required parameters, we use the delay equation itself,
Eq.(\ref{eq:evol1}), 
\begin{equation}
\dot{d}=-ad+\chi m+D_{d}d_{ss}+\Lambda{\cal I}[d],\label{eq:delayeq}
\end{equation}
The regression design variables $X_{1,2,3,4}$ are defined as
\begin{equation}
X_{1}=-d,\quad X_{2}=m,\quad X_{3}=d_{ss},\quad X_{4}={\cal I}[d],\label{eq:regress}
\end{equation}
with the coefficient vector
\begin{equation}
\bm{c}=(a,\chi,D_{d},\Lambda)^{\mathsf{T}},
\end{equation}
along with positivity constraints
\begin{equation}
(a,\chi,Dd,\Lambda)\geq0.\label{eq:pos}
\end{equation}

\subsection{Algorithm}

The algorithm for finding the reduced coefficients is as follows:
\begin{enumerate}
\item Choose the sets for $(\beta_{m},\gamma_{m},\tau_{m})$ as given in
Eqs.(\ref{eq:set1}-\ref{eq:set3}).
\item Reconstruct a memory field for each triplet $(\beta_{m},\gamma_{m},\tau_{m})$
using Eq.(\ref{eq:mem}).
\item Calculate the character-equilibrium constraint $R_{c}$ using Eq.(\ref{eq:char-eqlb})
\item Perform a regression using these values with regression design variables
as $X_{1,2,3,4}$, defined in Eq.(\ref{eq:regress}) along with imposition
of the positivity constraint (\ref{eq:pos}).
\item Calculate $a$ using the constraint (\ref{eq:constraint}) on $a$.
\end{enumerate}
Once we have all the values, we calculate the root mean-square error
\begin{equation}
\textrm{RMSE}=\sqrt{\frac{1}{N_{\textrm{test}}}\sum_{q}\left(\dot{d}^{\mathrm{num}}_{q}-\dot{d}^{\mathrm{model}}_{q}\right)^{2}},
\end{equation}
where $\dot{d}^{\mathrm{num}}_{q}$ and $\dot{d}^{\mathrm{model}}_{q}$
refer to the estimates of the rate of change of the chorus-character
delay at the same regression observation $q$, with $\dot{d}^{\mathrm{num}}_{q}$
calculated directly from consecutive delay-map entries and $\dot{d}^{\mathrm{model}}_{q}$
predicted by the fitted two-field reduced equation. The index $q$
is a label for the $i$th oscillator and $n$th chorus pair
\begin{equation}
q\longleftrightarrow(i,n).
\end{equation}
The numerical derivative is calculated from the relation
\begin{equation}
\dot{d}^{\mathrm{num}}_{q}=\frac{d^{(n+1)}_{i}-d^{(n)}_{i}}{\Delta t_{n}},
\end{equation}
 and the model-dependent derivative is predicted from the delay equation
Eq.(\ref{eq:delayeq})
\begin{eqnarray}
\dot{d}^{\mathrm{model}}_{q} & = & -ad_{q}+\chi m_{q}+D_{d}L_{q}+\Lambda H_{q},\\
L_{q} & = & d^{(n)}_{i-1}-2d^{(n)}_{i}+d^{(n)}_{i+1},\\
H_{q} & = & \sum_{j\in{\cal N}_{i}}w_{ij}\left[d_{j}(t^{(n)}_{c}-\tau_{ij})-d^{(n)}_{i}\right],
\end{eqnarray}
where ${\cal N}_{i}$ is the neighbour set for oscillator $i$. The
weights $w_{ij}$ are determined with the assumption that every neighbour
is assigned equal influence. Thus
\begin{equation}
\begin{array}{rcl}
w_{ij} & = & \frac{1}{2P},\quad j\in{\cal N}_{i},\\
 & = & 0,\quad\textrm{otherwise}.
\end{array}
\end{equation}
Also, the weights satisfy the normalisation condition
\begin{equation}
\sum_{j\in{\cal N}_{i}}w_{ij}=1.
\end{equation}

The best fitted values and the grid-selected values are determined
by the smallest RMSE, which are found to be
\begin{eqnarray}
\chi & = & 0.00358,\\
D_{d} & = & 0.00149,\\
\Lambda & = & 0.0329.
\end{eqnarray}
For the best fitted values, the value of $a$ is given by
\begin{equation}
a=\chi R_{c}=0.00398.
\end{equation}
The corresponding triplet values are 
\begin{equation}
(\beta_{m},\gamma_{m},\tau_{m})=(1,2,17).
\end{equation}

\end{appendices}

\bibliography{chimera-paper-sr-references}

\end{document}